\documentclass[proceedings]{JHEP3}
\PrHEP{PrHEP unesp2002}                 
\conference{Workshop on Integrable Theories, Solitons and Duality}

\def\rf#1{(\ref{eq:#1})}
\def\lab#1{\label{eq:#1}}
\def\nonu{\nonumber}
\def\br{\begin{eqnarray}}
\def\er{\end{eqnarray}}
\def\be{\begin{equation}}
\def\ee{\end{equation}}
\def\lb{\lbrack}
\def\rb{\rbrack}

\def\({\left(}
\def\){\right)}
\def\v{\vert}                     

\def\sskp{\par\vskip 0.15cm \par\noindent}
\def\bc{\begin{center}}
\def\ec{\end{center}}

\def\tr{\mathop{\rm tr}}                  
\def\Tr{\mathop{\rm Tr}}                  
\newcommand{\dder}[2]{\frac{d{#1}}{d{#2}}}
\newcommand{\pder}[2]{\frac{\partial{#1}}{\partial{#2}}}
\newcommand\sbr[2]{\left\lbrack\,{#1}\, ,\,{#2}\,\right\rbrack} 

\def\a{\alpha}
\def\b{\beta}

\def\d{\delta}

\def\eps{\epsilon}

\def\g{\gamma}

\newcommand{\h}{\frac{1}{2}}
\def\l{\lambda}

\def\o{\over}
\def\om{\omega}

\def\p{\phi}

\def\pa{\partial}

\def\pr{\prime}

\def\t{\tau}

\def\ti{\tilde}

\def\cD{{\cal D}}

\def\cL{{\cal L}}

\def\cU{{\cal U}}

\font \msb=msbm10 scaled \magstep1

\newcommand{\IC}{\mbox{\msb C} }
\newcommand\threemat[9]{\left(\begin{array}{ccc}  
{#1} & {#2} & {#3} \\ {#4} & {#5} & {#6} \\
{#7} & {#8} & {#9} \end{array} \right)}
\newcommand\fourmat[4]{\left(\begin{array}{cc}  
{#1} & {#2} \\ {#3} & {#4} \end{array} \right)}

\newcommand{\ct}[1]{\cite{#1}}
\newcommand{\bi}[1]{\bibitem{#1}}

\title{WDVV Equations, Darboux-Egoroff Metric and the Dressing Method}
\author{\speaker{H. Aratyn}\\
Department of Physics,
University of Illinois at Chicago,
845 W. Taylor St.,
Chicago, IL 60607-7059  \\
\email{aratyn@uic.edu}}
\author{J.F. Gomes\\Instituto de F\'{\i}sica Te\'{o}rica-UNESP,
Rua Pamplona 145,
01405-900 S\~{a}o Paulo, Brazil\\
\email{jfg@ift.unesp.br}}
\author{J.W. van de Leur\\Mathematical Institute,
University of Utrecht,
P.O. Box 80010, 3508 TA Utrecht,
The Netherlands\\
\email{vdleur@math.uu.nl}}
\author{A.H. Zimerman\\Instituto de F\'{\i}sica Te\'{o}rica-UNESP,
Rua Pamplona 145,
01405-900 S\~{a}o Paulo, Brazil\\
\email{zimerman@ift.unesp.br}}
\abstract{ Dressing technique is used to construct commuting Lax operators which provide
 an integrable (canonical) structure behind Witten--Dijkgraaf--Verlinde--Verlinde
equations. 
The commuting flows are related to the isomonodromic flows.
Examples of the canonical integrable structure are given
in two- and three-dimensional cases.
The three-dimensional example is associated with 
the rational Landau-Ginzburg potentials.}

\begin{document}

\section{Introduction, WDVV Equation in Flat Coordinates}
\label{intro}
In this talk we 
describe the commuting structure behind Witten--Dijkgraaf--Verlinde--Verlinde
(WDVV) associativity equations based on the dressing approach.
The solutions to the 
WDVV equations \ct{witten,dvv,Du-lectures,Du3a}, 
expressed in  terms of the so-called flat coordinates
$x^1, x^2, {\ldots} ,x^N$,
are provided by the prepotential
$ F (x^1, x^2, {\ldots} ,x^N)$ and
the Euler operator  :
\be 
E = \sum_{\a=1}^N \( d_\a x^\a + r_\a \) \frac{\pa}{\pa x^\a}
\lab{eflat}
\ee
with $d_\a r_\a =0 $ and $d_\a= 1+\mu_1 -\mu_\a$ \ct{Du-lectures},
with constants $\mu_\a, \a=1,{\ldots} ,N$ defined below (see
\rf{mudef}).

The pair $(F,E)$ satisfies the \underbar{{WDVV equations}} if
the following three conditions are satisfied.
  
$\bullet$   \underbar{{Associativity}} :
$$
\sum_{\d,\g=1}^N \frac{\pa^3 F (x) }{\pa x^\a \pa x^\b \pa x^\d }\eta^{\d \g}
\frac{\pa^3 F (x) }{\pa x^\g \pa x^\omega \pa x^\rho }
=\sum_{\d, \g=1}^N
\frac{\pa^3 F (x) }{\pa x^\a \pa x^\omega \pa x^\d }
\eta^{\d \g}
\frac{\pa^3 F (x) }{\pa x^\g \pa x^\b \pa x^\rho }
$$

$\bullet$   \underbar{{Normalization}} :
$$  
\frac{\pa^3 F (x) }{\pa x^\a \pa x^\b \pa x^1 }
= \eta_{\a\b} 
$$
where $\eta_{\a\b}$ defines a \underbar{constant} non-degenerate metric:
$g = \sum_{\a \b =1}^N \eta_{\a \b} dx^\a dx^\b$.

$\bullet$   \underbar{{Quasi-Homogeneity}} condition :
The quasi-homogeneity condition states that :
\be
E (F) = d_F F + \, {\rm quadratic~terms}
\lab{quasi}
\ee
where the number $d_F$ denotes the degree (or homogeneity) 
of the prepotential $F$.

As an example consider the three-dimensional space 
with three flat coordinates $x^1,x^2,x^3$ and with $(E,F)$ :
\br
F &= &\frac{1}{6} x^3 (x^2)^3 + \frac{1}{6} (x^1)^3 + x^1 x^2 x^3
+ \h (x^3)^2 \( \log x^3 - \frac{3}{2}\) \nonu\\
E  &= & x^1 \frac{\pa}{\pa x^1}+ \frac{1}{2} x^2 \frac{\pa}{\pa x^2}
+ \frac{3}{2} x^3 \frac{\pa}{\pa x^3}  \nonu
\er
such that $E (F) = 3 F + \, {\rm quadratic~terms}$.
More details on this example will be given in 
Section \ref{lg-models}.

The content of this talk is as follows.
In Section \ref{de-metric}, we define  
the Darboux-Egoroff equations which characterize the 
metric behind WDVV solutions when expressed in terms
of the curvlinear orthogonal coordinates referred to as canonical coordinates.
The canonical integrable structure behind the WDVV equations is 
presented in Section \ref{integra} emphasizing its connection 
to the Darboux-Egoroff metric. We use the setting of the
Riemann-Hilbert problem augmented by an extra twisting condition
\ct{vandeLeur:2000gk}.
The tau function appears naturally in this formalism.
The dressing matrix entering the Riemann-Hilbert problem
generates the dressing procedure which is used to construct the commuting
structure behind the WDVV equations.
The dressing procedure is developed in Section \ref{dressing}
and used to provide relation between the canonical integrable structure
and the flat coordinates, structure
constants and associativity equations.
Section \ref{monodromy} establishes a connection between 
the commuting flows of the canonical integrable structure
behind the WDVV equations and isomodromic deformations
related to the Schlesinger equation.
Another evidence of such connection is provided by the fact that the 
tau function of the Riemann-Hilbert problem turns into the isomonodromic
tau function once the conformal condition on the
integrable structure is imposed as explained in Section \ref{dressing}
(see \ct{LM,needs-paper}).

In case of two-dimensions, solutions to the Darboux-Egoroff
equations, the tau functions and the corresponding prepotential
satisfying WDVV equations can be found explicitly. This is described 
in Section \ref{ntwo}.
More difficult is the case of three dimensions presented in Section 
\ref{nthree} where the Darboux-Egoroff equations are shown to take 
the form of the  classical Euler 
equations of free rotations of a rigid body.
In three-dimensions the scaling dimension of the tau function is found
to be related to the integral of the  Euler 
equations.

Section \ref{lg-models} shows how to derive the canonical 
integrable structures
for a class of rational Lax functions associated
with a particular reduction of the dispersionless 
KP hierarchy. This derivation generalizes the well-known
construction of the monic polynomials \ct{Du-lectures,Hlectures}.
An example  of the three-dimensional canonical integrable
model derived from the rational potentials is given
in subsection \ref{examples}.
The three-dimensional example shown in this subsection 
provides solutions to the Painlev\'e VI equation. 
Given that the flows of the canonical integrable models
can essentially be reformulated as isomonodromic deformations,
as shown in Section \ref{monodromy}, the connection to the
sixth Painlev\'e equation is not surprising.
The tau function of the three-dimensional example has
a scaling dimension of $R^2=1/4$ and the corresponding prepotential
contains logarithmic terms.

For the scaling dimensions, $R^2=n^2$ such that $n$ is an integer,
the multi-component KP hierarchy provides a framework
for the construction of canonical integrable hierarchies
\ct{prepare}.
It would be of interest to find a universal approach 
to the formulation of the canonical integrable models 
which would include models with fractional scaling dimensions
as the ones encountered in example of subsection \ref{examples}
based on the rational potentials of Section \ref{lg-models}.

\section{Darboux--Egoroff metric }
\label{de-metric}
Massive topological field theories can be  classified locally by the
Darboux--Egoroff metric given in terms of the canonical coordinates
$u_1, {\ldots} ,u_N$ \ct{Du-lectures} :
\be
g = \sum_{\a \b =1}^N \eta_{\a \b} dx^\a dx^\b= 
\sum_{i=1}^N h_i^2 ({\bf u}) (d u_i)^2
\lab{lam}
\ee
with  Lam\'e coefficients $ h_i^2 ({\bf u}) = \pa \phi / \pa u_i$.
The fact that $h_i^2(u)$ is a gradient ensures that the 
so-called ``rotation coefficients''
\be
\b_{ij} = \frac{1}{h_j} \frac{\pa h_i}{\pa u_j}, \;\;  i\ne j,
\;\; 1\le i,j\le N ,
\lab{rotco}
\ee
are symmetric $\b_{ij} = \b_{ji}$ and therefore the metric becomes
the Darboux-Egoroff metric when expressed in terms of the 
curvlinear orthogonal or canonical coordinates $u_i$.
The Darboux-Egoroff equations for the rotation coefficients are:
\begin{equation}
    \frac{\partial }{ \partial u_k} \beta_{ij} = \beta_{ik} \beta_{kj},
\;\; \;\mbox{distinct}\;\; i,j,k
\lab{betas-comp}
\end{equation}
\be
\sum_{k=1}^{N} \frac{\partial }{ \partial u_k} \beta_{ij}     =0,
\;\; i \ne j \, .
\lab{ionb}
\ee
In addition one also assumes the conformal condition :
\begin{equation}
\sum_{k=1}^Nu_k\frac{\partial }{ \partial u_k} \beta_{ij} = -\beta_{ij}\, .
\lab{betas-deg}
\end{equation}

\section{Integrable structure behind WDVV equations}
\label{integra}
Consider a loop group element $g(z ) : S^1 \to  {GL} (N, \IC)$
which decomposes as 
$$
g(z ) =  g_{-} (z ) g_{+}(z ) 
$$
w.r.t. two subgroups of the Lie loop group 
$G= \widehat{GL} (N, \IC)$ consisting of all such maps $g$:
\br
G_{-} &=& \left\{ g \in G \v \; g(z) = 1 + \sum_{i>0} g^{(-i)} z^{-i}  \right\}
\nonu \\
G_{+} &=& \left\{ g \in G \v \; g(z) = \sum_{i\geq0} g^{(i)} z^i  \right\}
\nonu
\er
Assume from now on that $g(z)$ satisfies twisting condition  
$g^{-1}(z)=g^T(-z)$ \ct{vandeLeur:2000gk}.
Let the un-dressed wave matrix be :
\br
\Psi_0 ({\bf u}, z)  & =&  
\exp\left({\sum_{j=1}^{N} z E_{jj}u_j}\right)
= \exp\left( z U \right) \lab{gms} \\
U & =& {\rm diag} \( u_1, {\ldots} ,u_N\)=\sum_{i=1}^N
 u_i E_{ii} \, .
\nonu
\er
Our notation is : 
$({\bf u})= ({ u_1}, {\ldots} , { u_{N}})$,
$\partial_j = {\partial}/ {\partial u_j}$ and matrix
$E_{ij}$ has matrix elements
$\( E_{ij}\)_{kl}= \d_{ik} \d_{jl}$.

The Riemann-Hilbert problem is here defined as :
\be
\Psi_0 ({\bf u}, z) \, g (z) \\
= \Theta^{-1} ({\bf u},z) \,  M  ({\bf u},z) 
\lab{rh-def}
\ee
with the dressing matrices :
\be
\Theta ( {\bf u} , z) \in G_{-} , \;\;\;
\Theta = 1 + \theta^{(-1)} z^{-1} + \theta^{(-2)} z^{-2}+{\ldots} 
\lab{grad-exp}
\ee
\be
M ( {\bf u} , z) \in G_{+} , \;\;\;
M = M_0 +M_1 z +M_2 z^2 + {\ldots} 
\nonu
\ee
satisfying the twisting conditions \ct{vandeLeur:2000gk,agz01}
\be
\Theta^{-1} ( {\bf u}, z )=     \Theta^T ( {\bf u}, -z )
\, , \qquad \;\;
M^{-1} ( {\bf u}, z )=    M^T ( {\bf u}, -z ) \,. 
\lab{red-theta}
\ee
The twisting conditions  imply, in particular, that 
$$
M_0^T = M_0^{-1} \quad ;\quad
\theta^{(-1)} = \theta^{(-1)\, T}
$$

The Riemann-Hilbert problem \rf{rh-def} gives rise to
the commuting 
symmetry flows :
\br
\frac{\partial}{\partial u_j} \Theta ({\bf u}, z) 
&=& - \left(\Theta z E_{jj} \Theta^{-1} 
\right)_{-} \Theta ({\bf u}, z) 
\lab{uthpos}\\
\frac{\partial}{\partial u_j} M ({\bf u}, z) 
&=&  \left(\Theta z E_{jj} \Theta^{-1} 
\right)_{+} M ({\bf u}, z) \,. 
\lab{ummpos}
\er
Equations \rf{uthpos} imply that the following tracelessness condition
\be
I \( \Theta ({\bf u}, z)\) =\sum_{j=1}^{N} \frac{\partial}{\partial u_j} \, 
\Theta ({\bf u}, z) =0 
\lab{traceum}
\ee
 holds for the so-called identity vector-field 
\be
I= \sum_{j=1}^N \pa /{\pa u_j}\, .
\lab{id-vect}
\ee
We also define the so-called Euler vector field in terms of 
canonical coordinates as 
\be
E =  \sum_{i=1}^N u_i \pder{}{u_i}\, .
\lab{euler-can}
\ee
One finds from equations \rf{uthpos} that
\be
E (\Theta )\Theta^{-1} = - \left(\Theta z U \Theta^{-1} 
\right)_{-} \,. 
\lab{eonth}
\ee
The tau function can be associated with 
the Riemann-Hilbert problem \rf{rh-def} through relations :
\be
\frac{\pa  \log \tau }{\pa u_j} =
{\rm Res}_{z} \(\tr \(  E_{jj} {\Theta}^{-1}  
z \dder{\Theta}{ z} \)  \) , \; \, j=1, {\ldots} ,N \,. 
\lab{jnj-tau}
\ee 
Accordingly, we introduce the following parametrization
for the symmetric $\theta^{(-1)}$ matrix :
\be
\theta^{(-1)}_{ij}= \left\{ \begin{array}{cc} 
- {\pa  \log \tau }/{\pa u_i} & \;\;\; i=j \\
\b_{ij} & \;\;\; i \ne j 
\end{array}  \right.
\lab{them1}
\ee
with off-diagonal elements of $\theta^{(-1)}$-matrix defining the  
so-called rotation coefficients $\b_{ij}$ satisfying 
the Darboux-Egoroff equations \rf{betas-comp} and
\rf{ionb} as follows from expressions \rf{uthpos} and \rf{traceum}.

\section{Dressing}
\label{dressing}
The un-dressed structure is given by operators :
\be
\d_j = {\pa  \o \pa u_j}  - z E_{jj}
, \quad\;\; j = 1, {\ldots} , N
\lab{delnj}
\ee
and
\be
L_k =   - z^{k+1} \frac{d}{d z} +  
 U z^{k+1} , 
\lab{viraun}
\ee
which both annihilate the un-dressed wave function 
$\Psi_0 ({\bf u}, z) $ \rf{gms} 
\be
 \d_j \Psi_0 ({\bf u}, z) = 0 , \quad
 L_k \Psi_0 ({\bf u}, z)= 0  \,. 
\lab{anni}
\ee
These operators satisfy the commutation relations :
\be
\lbrack \d_j \, , \, \d_i \rbrack =0,\;\;\;
\lbrack {L_k}, {L_r}\rbrack = (k-r) L_{k+r}
, \;\;\; \lbrack {L_k}, \d_j\rbrack = 0 \,. 
\lab{comma} 
\ee
The so called {dressing procedure} maps the un-dressed wave function 
$\Psi_0 ({\bf u}, z) $ to :
$$
\Psi_0 ({\bf u}, z) \; \to \;
\Psi  ({\bf u}, z) =\Theta ({\bf u}, z) \Psi_0 ({\bf u}, z)
$$
while the operators  $\d_j$ and $ L_k$ are mapped 
into the dressed operators:
\be
\d_j \; \to \; \cD_j = \Theta \d_j \Theta^{-1}, \quad
L_k \; \to \; \cL_k = \Theta L_k \Theta^{-1} \,.
\lab{dresop}
\ee
By construction these operators annihilate the ``dressed'' wave (matrix) function
$\Psi  ({\bf u}, z)$ : 
\be
\cL_k \Psi = 0  , \quad \cD_j \Psi = 0  , \quad j=1, {\ldots} , N\,.
\lab{annid}
\ee
Dressing preserves the commutation relations and so it holds that
$\lbrack \cD_j \, , \, \cD_i \rbrack=0$
and $\lbrack {\cL_k}, {\cL_r}\rbrack = (k-r) \cL_{k+r}$
, $\lbrack {\cL_k}, \cD_j\rbrack = 0$.

Since
\be
\Theta  \pa_j \Theta^{-1} = \pa_j + \Theta ( \pa_j \Theta^{-1} )
= \pa_j + \( \Theta z E_{jj} \Theta^{-1} \)_{-}
\lab{tpaiti}
\ee
we obtain 
\be
\cD_j =  \pa_j -  \( \Theta z E_{jj} \Theta^{-1} \)_{+}
=\pa_j - z E_{jj} - V_j \,. 
\lab{tpaitia}
\ee
Also,
\be
\cL_{0} = \Theta L_{0} \Theta^{-1}=
- z\frac{d}{d z} + z U +  V +\( z\frac{d \Theta}{d z}-
E (\Theta )\) \Theta^{-1} 
\lab{lmnsor}
\ee
with
\br
V_j &\equiv& \lbrack \theta^{(-1)} \, , \, E_{jj}\rbrack , \;\;\;
(V_j)_{kl}  = \( \d_{lj} -  \d_{kj}\)  \b_{kl} 
\lab{defvi} \\
V &\equiv& \lbrack \theta^{(-1)} \, , \, U \rbrack, \;\;\;
V_{ij} =  (u_j-u_i) \beta_{ij} \,.  
\lab{v-comps}
\er
Note, that the first three terms on the right hand side of equation
\rf{lmnsor} contain only terms of positive grade in $z$ while 
the remaining term contains terms of the negative grade in
$z$ (terms with $z^k$ , $k \le -1$).
We will now impose the so-called conformal condition which amounts
to $(\cL_{0})_+ = \cL_{0}$ or
\be
 E (\Theta)  = z \frac{d \Theta}{d z} \, .
\lab{conformalc}
\ee
As shown in reference \ct{needs-paper}
this condition is compatible with flows from equation 
\rf{uthpos}

The following equations follow now from relations \rf{annid}
\br
\cD_j \Psi &=& 0 \;\to \; \frac{\pa \Psi}{\pa u_j}= (V_j + z E_{jj}) \Psi
\lab{djpsi} \\
\cL_{0} \Psi &=& 0  \;\to \; z \frac{d \Psi}{d z}= (z U + V) \,. 
\Psi \lab{vironpsi}
\er
{}From \rf{djpsi} one can easily calculate the action of the Euler
vector field on the wave-function $\Psi$:
\be
E (\Psi) = \sum_{j=1}^N u_j  \frac{\pa \Psi}{\pa u_j} = 
(V +z U ) \Psi
\lab{eulerpsi}
\ee
as follows from relations $ \sum_{j=1}^N u_j V_j =V$ and 
$\sum_{j=1}^N u_j E_{jj}=U$.
Comparing this with equation \rf{vironpsi} 
we find that :
\be
 E (\Psi)  = z \frac{d \Psi}{d z}  
\lab{confcond}
\ee
and so $E$ and $z d /d z$ coincide
when applied on the wave function $\Psi$ in the framework of the dressing formalism.
Plugging relation \rf{conformalc} into the formula
\rf{jnj-tau} for the $\t$-function one obtains \ct{Aratyn:2001cj}:
\be
\pa_j \log \t = 
{\rm Res}_{z} \(\tr \(  {\Theta}^{-1}  
E (\Theta) E_{jj}   \)  \) = \h \tr \( V_j V\) \,. 
\lab{tauvjv}
\ee 

Also, $I (\Psi) = z \Psi $, 
which together with relation \rf{conformalc} yield
\br
E (\b_{ij})&=&-\b_{ij}\; ;  \quad \quad
\; \; \; I ( \b_{ij})  =0
\lab{ebib} \\
E (\tau) &=& R^2 \; \tau \; ;  \quad   
I ( \tau) =0
\lab{ilogtv}
\er
where in relation \rf{ilogtv} we introduced constant $R^2$  defining
the scaling dimension (also called homogeneity) of the tau function.

Commutation relations yield :
\be
\pa_j V = \lbrack V_j \, , \, V \rbrack , \;\;\quad
 \lbrack V  \, , \, E_{jj}\rbrack = \lbrack V_j  \, , \, U\rbrack
\lab{pjv}
\ee
and so $I(V)=0$ since $\sum_{j=1}^N V_j=0$.

The similarity transformation $V \to 
{\cal V} = M_0^{-1} V M_0 $ transforms $V$ to the constant matrix
${\cal V}$ ($\pa_j {\cal V} =0$) due to the flow equations 
$\pa_j M_0= V_j M_0$, which follow from relation \rf{ummpos},
and the above equation \rf{pjv}.
Assume, now that there exists an invertible 
matrix $S$ which diagonalizes ${\cal V}$ \ct{Du-lectures}:
\be
S^{-1} {\cal V}S=\mu = \sum_{j=1}^N \mu_j E_{jj} 
\lab{mudef}
\ee
where $\mu$ is a constant diagonal matrix $ \mu = {\rm {diag}}
 (\mu_1, {\ldots} , \mu_N)$.

Next, define a matrix
\be
M(u)=M_0(u)S=(m_{ij}(u))_{1\le i,j\le N}\, .
\lab{mdef}
\ee
$M$ satisfies 
\be
(\pa_j - V_j) (M)=0 \;\; \to \;\; M^{-1} (\pa_j M )= M^{-1} V_j M
\lab{msatis}
\ee
and  
\be
E (M)  = \sum_{j=1}^N u_j V_j M=V M = M_0 {\cal V}S=   M \mu \, .
\lab{euleronm}
\ee
Define a constant non-degenerate metric to be  :
\be
\eta=(\eta_{\alpha\beta})_{1\le \alpha,\beta\le N}=M^TM=S^TS,
\quad \hbox{and denote}\quad \eta^{-1}=(\eta^{\alpha\beta})_{1\le
\alpha,\beta\le N} \, .
\ee
Hence $\eta_{\a \b} 
= \sum_{i=1}^N m_{i\a} m_{i\b}$. 

We have :
\be
( \frac{\pa}{\pa u_j}- V_j -z E_{jj}) \Phi = 0 
\lab{cdjphi}
\ee
where 
\be
\Phi ({\bf u}, z) =  \Psi ({\bf u}, z)  g(z) S =
M ({\bf u}, z) S= M({\bf u}) + O(z)  \,. 
\lab{phi-def}
\ee
Under a similarity transformation generated by the $M(u)$ matrix
the Lax operators $\cD_j =\pa_j - z E_{jj} - V_j $ transform to:
\be
{\widetilde \cD}_j \equiv M^{-1} \cD_j M
= \pa_j - z  M^{-1} E_{jj} M
= \pa_j - z C_j \, , 
\lab{wcdj}
\ee
where in \rf{wcdj} we have introduced :
\be
C_j = M^{-1} E_{jj} M = \sum_{\a=1,\b=1}^N (C_j)_{\b}^\a E_{\a \b},  \;\;\;
(C_j)_{\b}^\a = \sum_{\g }\eta^{\a \g} m_{j \g} m_{j \b}
\lab{defcj}
\ee
such that 
\be
C_i C_j= C_i \d_{ij}, \quad  \sum_{i=1}^N C_i =I_N  \,.  
\lab{cicj}
\ee

The matrix $\Xi (u,z) $ defined as 
\be
\Xi (u,z) = M^{-1} \Phi(u,z)  = I + \sum_{n=1}^{\infty} z^n 
\Xi^{(n)} (u)= 
I + z \Xi^{(1)} + z^2 \Xi^{(2)}+ \cdots
\lab{minpexpc}
\ee
is annihilated by the transformed ${\widetilde \cD}_j$ operators:
\be
{\widetilde \cD}_j ( \Xi)= (\pa_j - z C_j) ( \Xi)= 
0 \,.
\lab{cdannimp}
\ee

Due to $\sbr{{\widetilde \cD}_i}{{\widetilde \cD}_j}=0$
it holds that:
\be
\pa_i C_j -\pa_j C_i= 0  \,. 
\lab{matrc}
\ee
{}From \rf{matrc}, it follows that we can define a matrix $C$ 
\be
C = \sum_{\a=1,\b=1}^N C_{\b}^\a E_{\a \b} 
\lab{cmdef}
\ee
 such that :
 \be
 C_j = \pa_j C  \,. 
 \lab{cdef}
 \ee
Accordingly, plugging the expansion \rf{minpexpc}
into equation \rf{cdannimp} yields :
\be
\pa_j \Xi^{(n)}  = C_j \Xi^{(n-1)} =\( \pa_j C \)\, \Xi^{(n-1)}, 
\;\;\; n \ge 1, \;\; \Xi^{(0)}=I
\lab{eqsxi}
\ee
and therefore we can choose $C$ to be equal to $\Xi^{(1)}$:
\be 
\Xi^{(1)} = C 
\lab{xionc}
\ee
while $ \pa_j \Xi^{(2)}  =  (\pa_j C) C$, etc.
Also, by summing over $j$ in \rf{eqsxi} we get
\be
\sum_{j=1}^N \pa_j \Xi^{(n)}  =  \Xi^{(n-1)} \;\; {\rm or}\;\;
\sum_{j=1}^N \pa_j \Xi= I (\Xi)  = z \Xi  \,. 
\lab{sumjxi}
\ee
Note, that the relation \rf{confcond} yields :
\be
z \frac{ d }{d z } \Xi = z \cU \Xi + \sbr{\mu}{ \Xi} \,. 
\lab{confcon}
\ee
In components it gives
\be
n  \Xi^{(n)} - \sbr{\mu}{\Xi^{(n)}} = \cU \Xi^{(n-1)} \,. 
\lab{confconc}
\ee
Comparing with \rf{eeqsxi} we see that
$E = n I - {\rm ad}_{\mu} $ when applied on $\Xi^{(n)}$.
Equation \rf{eqsxi} determines $\Xi^{(n)} $  recursively up to a constant.
Note, that we can add to $\Xi^{(n)} $ a constant $A_n$ 
such that $nA_n - \sbr{\mu}{A_n}=0$ without changing the 
right hand side of \rf{confconc}.
Hence, $\Xi^{(n)} $ which is a solution of equation \rf{confconc} 
as well equation \rf{eqsxi} 
is determined up to a constant 
$A_n$ of grade $n$ with respect to the 
the grading defined by the semisimple element $\mu$ according to 
equation $nA_n - \sbr{\mu}{A_n}=0$.

The above ambiguity amounts to 
the fact that $\Xi$ is determined upto a
constant power series $A(z)$ such that
\[
A(z)=I+A_1z +A_2z^2+\cdots, \qquad A(z)\eta A(-z)^T=\eta.
\]
with $A_n$ satisfying $nA_n - \sbr{\mu}{A_n}=0$.

The matrix $C$ is crucial for the whole
theory and we will now study its properties. 
According to \rf{defcj} its matrix elements 
$C_{\b}^\a $ must satisfy :
\be
\pa_j C_{\b}^\a = \sum_{\g }\eta^{\a \g} m_{j \g} m_{j \b}
\;\; \to \;\; \pa_j C_{\b\a} = m_{j\a} m_{j\b} 
\lab{cmm}
\ee
where $C_{\b \a} = \sum_{\g=1}^N C^\g_\b \eta_{\a \g}$.
Therefore from equation \rf{cmm}:
\be
C_{\a \b} = C_{\b \a} \quad {\rm or} \quad
 \sum_{\g=1}^N C^\g_{\a}\eta_{\b \g} 
= \sum_{\g=1}^N  C^\g_{\b}\eta_{\a\g} \,. 
\lab{symmc}
\ee

Next, define flat coordinates as the first column of the 
$C$ matrix :
\be
x^\a \equiv {C_{1}^\a}\,.  
\lab{xadef}
\ee
The Jacobian of the change of variables from $u_i$ to
$x^\a$ is according to \rf{defcj} given by :
\be
\frac{\pa x^\a }{\pa u_i} = \sum_{ \b=1 }^N \eta^{\a \b} m_{i1}m_{i\b} \,. 
\lab{jac}
\ee
Also,
\be
\frac{\pa u_i}{\pa x^\a} = \frac{m_{i\a}}{m_{i1}}
\lab{jaca}
\ee
as follows from the identity: $\d_{ij} = \sum_{\a \b} m_{i\a} 
\eta^{\a \b}   m_{j\beta}$.

In terms of $x^\a$-coordinates the Lax operators
${\widetilde \cD}_j$  become:
\be
{\widetilde \cD}_\a= \sum_{j=1}^N \frac{\pa u_j}{\pa x^a} {\widetilde \cD}_j
= \frac{\pa }{\pa x^\a} - z C_\a \,. 
\lab{xlax}
\ee
The commutation relations :
\be
\left\lbrack \frac{\pa }{\pa x^\a} - z C_\a \, , \, 
\frac{\pa }{\pa x^\b} - z C_\b \right\rbrack =0
\lab{wdvv}
\ee
yield associativity 
\be
\left\lbrack C_\a \, , \, C_\b \right\rbrack =0 \; \to \;
\sum_{\g=1}^N \( c^{\g}_{\a \d} c^{\omega}_{\b \g} -
c^{\g}_{\b \d} c^{\omega}_{\a \g}\) =0
\lab{wdvva}
\ee
and integrability 
\be
\frac{\pa }{\pa x^\a} C_\b - \frac{\pa }{\pa x^\b} C_\a =0
\lab{ddc}
\ee
relations.
The matrix $C_\a$ :
\be
C_\a \equiv
\sum_{\g=1 \b=1}^N c^\g_{\a \b} E_{\g \b}
\lab{cadef}
\ee
is equal to 
\be
C_\a =\sum_{j=1}^N \frac{\pa u_j}{\pa x^a} C_j = \frac{\pa }{\pa x^a} C
\lab{intrc}
\ee
The structure constant $c_{\a \b \g}= 
\sum_{\d =1}^N \eta_{\a \d} c^\d_{\b \g}$ becomes therefore
equal to
\be
c_{\a \b \g} = \frac{\pa C_{\b \g}}{\pa x^\a} 
\lab{destrcon}
\ee
which according to \rf{jac} and \rf{cmm} is equal to :
\be
c_{\a \b \g} = \frac{\pa C_{\b \g}}{\pa x^\a} = \sum_{j=1}^N \frac{m_{j\a}}{m_{j1}}
\pa_j C_{\b\g} = \sum_{j=1}^N \frac{m_{j\a}m_{j\b} m_{j\g} }{m_{j1}} \,. 
\lab{sympxc}
\ee
Hence the structure constants $c_{\a \b \g}$ 
are symmetric in all three indices. 
Also, $c_{1 \b \g} = \eta_{\b \g}$.

It follows from \rf{sympxc} that 
\be
\frac{\pa C_{\b \g}}{\pa x^\a} = \frac{\pa C_{\b \a}}{\pa x^\g} 
\lab{xcxc}
\ee
and therefore 
\be
C_{\a \b} = \frac{\pa^2 F}{\pa x^\a \pa x^\b} \;\;\; {\rm
or} \;\;\;
c_{\a \b \g} = \frac{\pa^3 F}{\pa x^\a \pa x^\b \pa x^\g} 
\lab{cfxx}
\ee
where $F$ is called the prepotential.

Let us go back to the linear problem :
\be
\( \frac{\pa }{\pa x^\a} - z C_\a \) (M^{-1}  \Phi) =\( \frac{\pa }{\pa
x^\a} - z C_\a \) (\Xi) = 0 \,. 
\lab{xcmp}
\ee
Introduce
\be
\phi_\b \equiv \sum_{\a=1}^N \eta_{1 \a} (\Xi)^\a_\b 
= I + z \phi_\b^{(1)} + z^2 \phi_\b^{(2)} + z^3 \phi_\b^{(3)} +{\ldots} 
\lab{pbexp}
\ee
then as we will show below, the prepotential
is given by a closed expression :
\be
F = - \h \p_{1}^{(3)} + 
\h \sum_{\d=1}^N x^\d \p_{\d}^{(2)} \, .
\lab{fep}
\ee
To show this, multiply equation \rf{xcmp} by $\sum_{\a=1}^N \eta_{1 \a}$
which yields :
\be
\frac{\pa \phi_\b}{\pa x^\a} - z \( \eta M^{-1} \Phi\)_{\a\b}=0
\lab{paapb}
\ee
where we introduced:
\be
\phi_\b \equiv \sum_{\a=1}^N \eta_{1 \a} (M^{-1} \Phi )^\a_\b 
= \sum_{\a=1}^N{} \eta_{1 \a} \sum_{\g, \d=1}^N{}
\eta^{\a \g} m_{\d \g} \Phi_{\d \b} = \sum_{\d=1}^N{}
m_{\d 1} \Phi_{\d \b}
\lab{phiab}
\ee
and where use was made of the identity :
\be
\sum_{\g=1}^N{} \eta_{1 \g} (C_\a )^\g_\b = c_{1 \a \b} = 
\eta_{\a \b}
\lab{coab}
\ee
leading to 
\be
\sum_{\g ,\d=1}^N{} \eta_{1 \g} (C_\a )^\g_\d (M^{-1} \Phi )^\d_\b
= (\eta M^{-1}  \Phi)_{\a\b}\,. 
\lab{coaba}
\ee
Applying $\pa / \pa x^\g$ on \rf{paapb} and using repeatedly
\rf{paapb} one gets :
\br
\frac{\pa^2 \phi_\b}{\pa x^\a x^\g} - z \( \eta z C_\g M^{-1} \Phi\)_{\a\b}=
\frac{\pa^2 \phi_\b}{\pa x^\a x^\g} - z^2 \( \eta  C_\g \eta^{-1} 
\eta M^{-1} \Phi\)_{\a\b}&=&
\nonu \\
\frac{\pa^2 \phi_\b}{\pa x^\a x^\g} - z \sum_{\d=1}^N
c^\d_{\a \g} \( z \eta M^{-1} \Phi\)_{\a\b}=
\frac{\pa^2 \phi_\b}{\pa x^\a x^\g} - z \sum_{\d=1}^N
c^\d_{\a \g} \frac{\pa \phi_\b}{\pa x^\d} =0
\lab{ppbaa}
\er
where we used $ \sum_{\d \omega=1}^N{} \eta_{\a \d} (C_\g)^\d_\omega \eta^{\omega
\b}= c^\b_{\a \g}$.

{}From \rf{minpexpc} we find that $\phi_\b$ as defined in \rf{phiab}
can be expanded in $z$ as follows :
\br
\phi_\b &= &\sum_{n=0}^\infty \phi^{(n)}_\b z^n = 
\phi^{(0)}_\b + z \phi^{(1)}_\b + z^2 \phi^{(2)}_\b 
+ z^3 \phi^{(3)}_\b + \cdots =
\sum_{\g=1}^N \eta_{1 \g} \( e^{z C}\)^\g_\b \nonu\\
&= &
\eta_{1 \b} + z \sum_{\a=1}^N \eta_{1 \a} C^\a_\b + 
z^2 \sum_{\g=1}^N{} \eta_{1 \g} (\Xi^{(2)})^\g_\b 
+  z^3 \sum_{\g=1}^N{} \eta_{1 \g} (\Xi^{(3)})^\g_\b
+ \cdots 
\lab{phiexp}
\er
where we used relation \rf{xionc}.

{}From that we can read:
\be
\phi^{(0)}_\b  = \eta_{1 \b}, \qquad 
\phi^{(1)}_\b = \sum_{\a=1}^N \eta_{1 \a} C^\a_\b = 
\sum_{\a=1}^N  \eta_{\b\a} x^\a
\lab{phizo}
\ee
where in the last equation we have used \rf{symmc} and \rf{xadef}.
Also,
\be
\phi^{(2)}_\b = \sum_{\g=1}^N{} \eta_{1 \g} (\Xi^{(2)})^\g_\b 
\;\;\;\;, \;\quad
\phi^{(3)}_\b = \sum_{\g=1}^N{} \eta_{1 \g} (\Xi^{(3)})^\g_\b \,. 
\lab{phitt}
\ee

Next, we have from \rf{paapb} :
\be
\Xi^\a_\b (u,z) = z^{-1} \sum_{\d=1}^N{} \eta^{\a \d } \frac{\pa \p_{\b} (u,z)}
{\pa x^\d} \,. 
\lab{expxiz}
\ee
Combining the above two equations we find
\be
C^\a_\b  (u)  = \sum_{\d=1}^N{}  \eta^{\a \d } \frac{\pa \p_{\b}^{(2)} (u)}
{\pa x^\d} \;\; \; \to \;\;
C_{\a\b } (u)  = \frac{\pa \p_{\b}^{(2)} (u)}{\pa x^\a}
 = \frac{\pa \p_{\a}^{(2)} (u)}{\pa x^\b} \,. 
\lab{expxizt}
\ee
Comparing with \rf{cfxx} we see that :
\be
\p_{\b}^{(2)} = \frac{\pa F }{ \pa x^\b} \,. 
\lab{ptwof}
\ee
Also, from \rf{minpexpc} it follows that 
\be
(\Xi^{(2)})^\a_\b (u) = 
\sum_{\d=1}^N{}  \eta^{\a \d } \frac{\pa \p_{\b}^{(3)} (u)} \,. 
{\pa x^\d} 
\lab{mioc}
\ee
The condition $\Xi (z)\eta^{-1} \Xi(-z)^T=\eta^{-1} $ holds due to the
twisting condition.
Expanding in $z$ gives on level of $z$ relation $C_{\a \b} =C_{\a\b}$
which is already known.
On next, $z^2$ level we get: $ (\Xi^{(2)})^T \eta + \eta \Xi^{(2)}
=C^T \eta C=0$ or
\be
\sum_{\g=1}^N{} \eta_{\b \g} (\Xi^{(2)})^\g_\b 
+ \sum_{\g=1}^N{} \eta_{\a \g} (\Xi^{(2)})^\g_\b = \sum_{\g=1}^N{} \eta_{\a \g} 
C^\g_\d C^\d_\b \,. 
\lab{theot}
\ee
For $\b=1$ one gets from \rf{mioc} and \rf{expxizt} that \rf{theot}
is equivalent to 
\be
\p_{\a}^{(2)} + \frac{\pa \p_{1}^{(3)} }
{\pa x^\a} = \sum_{\d=1}^N C^\d_1 \frac{\pa \p_{\a}^{(2)} }{\pa x^\d}=
\sum_{\d=1}^N x^\d \frac{\pa \p_{\a}^{(2)} }{\pa x^\d} \,. 
\lab{theota}
\ee
Hence as in \ct{AKrV} we find :
\be
\p_{\a}^{(2)} = \frac{\pa F }{\pa x^\a} = -\frac{\pa \p_{1}^{(3)} }
{\pa x^\a} +
\sum_{\d=1}^N x^\d \frac{\pa \p_{\a}^{(2)} }{\pa x^\d}
\;\;\; \to \;\;  F = - \h \p_{1}^{(3)} + 
\h \sum_{\d=1}^N x^\d \p_{\d}^{(2)} \,. 
\lab{fepa}
\ee

We will now derive an expression for the Euler vector field $E$ in terms of
the flat coordinates.

Define 
\be
\cU =  M^{-1} U M =\sum_{\a,\b=1}^N \cU_{\b}^\a E_{\a \b}
\lab{cudef}
\ee
and notice, that
\be
\cU = \sum_{i=1}^N u_i M^{-1} E_{ii} M = \sum_{i=1}^N u_i \pa_i C
= E (C) \,. 
\lab{ezcu}
\ee

{}From \rf{ezcu} and the fact that the first column of 
matrix $C$ defines the flat variables 
as in \rf{xadef} we find that
\be
E (x^\a) = E (C^\a_1) = \cU^\a_1
\lab{exaca}
\ee
from which follows an expression
for the Euler operator in terms of flat variables $x^\a$ 
 :
\be
E= \sum_\a \cU^\a_1 \pder{}{x^\a}\,. 
\lab{eulalter}
\ee

Furthermore, from \rf{eqsxi}
\be
E (\Xi^{(n)}) =  \sum_j u_j \pa_j \Xi^{(n)}  = \sum_j u_j 
C_j \Xi^{(n-1)} = \cU \Xi^{(n-1)}, 
\;\;\; n \ge 1, \;\; \Xi^{(0)}=I\,. 
\lab{eeqsxi}
\ee

Put $n=1$ in \rf{confconc},
it gives :
\be 
 \Xi^{(1)}- \sbr{\mu}{\Xi^{(1)}}=  C - \sbr{\mu}{  C} = \cU \,. 
\lab{pajcua}
\ee
Hence :
\be
\cU_{\b}^\a = (1-\mu_\a+\mu_\b) C_{\b}^\a
\lab{cuij}
\ee
and 
\be
\cU_1^{\a}= \( 1+ \mu_1 - \mu_\a \) x^\a \,. 
\lab{cu1axa}
\ee
Accordingly, the Euler vectorfield $E= \sum_i u_i \pa_i$ 
becomes in terms of the flat coordinates  
\be
E= \sum_{\a=1}^N \cU^\a_1 \pder{}{x^\a} =  \sum_{\a=1}^N  
(1+\mu_1-\mu_\a) x^\a \pder{}{x^\a} \,. 
\lab{eulaltera}
\ee
Similarly, for the identity vector filed $I$ we find
\be
I (x^\a )= \sum_{i=1}^n \frac{\pa x^\a }{\pa u_i} = \sum_{i, \b=1 }^N
\eta^{\a \b} m_{i1}m_{i\b} = \sum_{\b=1 }^N \eta^{\a \b} 
\eta_{1 \b} = \d_{\a 1}
\lab{jaci}
\ee
and therefore $I = \pa / \pa x^1$ in terms of the flat coordinates.

\section{Monodromy and Frobenius manifold}
\label{monodromy}
Let us first introduce notion of monodromy.
The notion of monodromy preserving deformations for linear differential
equations in the complex plane was first studied by Schlesinger \ct{sch12}.
Consider a linear differential equation with rational coefficients:
\be
{ d W \o dz } = A (z) \, W
\lab{diffeq}
\ee
where $A (z)$ is an $N \times N$ matrix valued function with rational entries.
In the case that  $A (z)$ has only simple poles in the finite plane one can
write:
\be
A (z) = \sum_{\nu=1}^N { A_{\nu} \o z - a_{\nu} }
\lab{poleexp}
\ee
In a neighborhood of any regular point for the differential eq. \rf{diffeq}
one can find a fundamental set of solutions $\{y_1 (z), \ldots, y_N (z) \}$.
If one analytically continues such a solution $y_j (z)$ around a singular
point $a_{\nu}$ it does not in general return to the solution  $y_j $
but to a linear combination $ \sum_k M^{\nu}_{kj}  y_k$.
The matrices $M^{\nu}$ are called monodromy matrices.

A question investigated by Schlesinger was as follows.
How must the coefficient 
matrices $A_{\nu}$ depend on the poles $\{a_1 , \ldots, a_n \}$ so that the
monodromy matrices $M^{\nu}$ do not depend  on the location of the poles.
The conditon for this takes a form of a non-linear system of equations
\be
\sum_\mu \pder{A_{\nu}}{a_\mu}  d a_{\mu} 
= - \sum_{\nu \ne \mu} \lb A_{\nu} \, , \, A_{\mu}\rb
{d a_{\nu}- d a_{\mu} \o a_{\nu}- a_{\mu} }\,. 
\lab{schleq}
\ee
These equations are known as Schlesinger equations.
In reference \ct{kyoto} it was shown that for solutions $A_{\nu} (a)$ to the 
Schlesinger equations \rf{schleq} the right hand side of this equation is
exact:
\be
\sum_\mu \pder{\log \( \tau (a) \)}{a_\mu}  d a_{\mu}  = \h \sum_{\nu \ne \mu} \Tr \( A_{\nu} A_{\mu}\)
{d a_{\nu}- d a_{\mu} \o a_{\nu}- a_{\mu} }\,. 
\lab{tau}
\ee
We will now show that that the canonical flows of the Frobenius 
manifolds reproduce
the structure of the Schlesinger equations.

Define
\be
S_i = M^{-1} E_{ii} ( V - \a I ) M
\lab{sidef}
\ee
where $\a$ is an arbitrary constant.

Recalling relations \rf{pjv} and \rf{msatis} one finds
\be
\pa_j S_i = M^{-1}\lbrack  E_{jj} \, , \, V_i \rbrack 
( V - \a I ) M  = \pa_i S_j 
\lab{pajsi}
\ee
due to the fact that $\lbrack  E_{jj} \, , \, V_i \rbrack 
=\lbrack  E_{ii} \, , \, V_j \rbrack$.

Thus, locally there exists a function $S$ such that $S_i = \pa_i S$.
A calculation based on relations \rf{pjv} and \rf{msatis} yields
\br
\frac{\partial}{\partial u_i} \ S_i &=&  \sum_{j=1,\ j\neq i}^N \ 
\frac{ \left[ S_i,S_j \right] }{u_i -u_j} \ ,\lab{paisi} \\
\frac{\partial}{\partial u_j} \ S_i &=&   
\frac{ \left[ S_i, S_j \right]}{u_j-u_i} \ ,\qquad \ i \neq j \,.  
\lab{pajsine}
\er
These are the Schlesinger equations. They can be rewritten in a 
more compact form as :
\be
d S_i = \sum_{j=1,\ j\neq i}^N  \left[ S_i,S_j \right]
\frac{d u_j- du_i}{u_j-u_i} , \quad \;\;\; d = \sum_j^N \pa_j d u_j\,.  
\lab{schlcomp}
\ee
It follows from the Schlesinger equations that
\be
S_{\infty}=  \sum_{j=1}^N S_i = M^{-1} ( V - \a I ) M =
{\cal V} - \a I
\lab{sinfty}
\ee
is constant as already established in \rf{mudef}.

The Schlesinger equations can be obtained as compatibility equations
of the following linear equations :
\be
\frac{d}{d \l} \ W = \ - \sum_{i=1}^N 
\frac{S_i}{\l-u_i} \ W  \ , \qquad
\pa_i  \ W =   \frac{S_i}{\l-u_i} \ W \,. 
\lab{wone} 
\ee
The compatibility equations reproduce 
equations \rf{paisi}-\rf{pajsine} by evaluating residues at 
$\{ u_i \}_{i=1,{\ldots} ,N}$

{}The system of
first-order differential equations \rf{wone} can be rewritten as
\be
\left( D \ +\ {\cal S} \right) W =0 \ ,
\lab{difeq}
\ee
where $D$ :
\be
D= \pder{}{\l} d \l + \sum_{i=1}^N
{\partial\over\partial u_i} d u_i 
\lab{extder}
\ee
is the exterior derivative 
and ${\cal S}$ the flat connection :
\be
{\cal S}= \sum_{j=1}^N\ S_j({\bf u}) \ D \log(\l-u_j)
= \sum_{j=1}^N\ S_j \frac{d \l - d u_j}{\l-u_j} 
\lab{conn}
\ee
which satisfies the zero-curvature condition :
\be
D\ {\cal S} +{\cal S} \wedge {\cal S} = 0 \ ,
\lab{cur}
\ee
due to \rf{difeq}.
Let, $ d U = \sum_{j=1}^N E_{jj} d u_j$.
One finds then that 
\be
D \log (\l - U) = 
(\l - U)^{-1} ( d \l - d U) =  \sum_{j=1}^N 
 \frac{E_{jj}}{\l-u_j} ( d \l - d U) 
\lab{tech}
\ee 
and the connection ${\cal S}$ can be rewritten as :
\be 
{\cal S} = M^{-1} (\l - U)^{-1} ( d \l - d U)  ( V - \a I ) M
= M^{-1} B M
\lab{sbm}
\ee
where in the last identity we defined
\be
B = (\l - U)^{-1} ( d \l - d U)  ( V - \a I )=\(D \log (\l - U) \)
( V - \a I )\,. 
\lab{bdef}
\ee
Let 
\be
{\cal D} = D + {\cal S} = D + M^{-1} B M
\lab{ddef}
\ee
then the zero-curvature condition \rf{cur} implies that
$ {\cal D}^2 =0$. 

It follows that
\be
{\widetilde {\cal D} } = M {\cal D} M^{-1} = D - d M M^{-1} +B
= D - A +B
\lab{dab}
\ee
with
\be
A = d M M^{-1} = \sum_{j=1}^N V_{j} d u_j
\lab{aadef}
\ee
will also define the flat connection, in agreement with 
\ct{Hlectures,boalch}.
In components :
\br
{\widetilde {\cal D} } &=&  \sum_{i=1}^N \( \pa_i - V_i - 
\frac{E_{ii}(V-\a I)}{\l-u_i}\) d u_i + \( \frac{\pa}{\pa \l}+ 
\sum_{i=1}^N  \frac{E_{ii}(V-\a I)}{\l-u_i}\) d\l \nonu \\
&=&
\sum_{i=1}^N \( \pa_i - V_i - \frac{E_{ii}(V-\a I)}{\l-U}\) d u_i 
+ \( \frac{\pa}{\pa \l}+ \frac{(V-\a I)}{\l-U}\) d\l \, .
\lab{tdcomp}
\er

Let $\Omega ({\bf u},\l)$ be such that the condition 
${\widetilde {\cal D} }\Omega ({\bf u},\l)  = 0$ holds.
In components this condition takes a form of
\br
\frac{\pa}{\pa \l} \Omega ({\bf u},\l) &=&    \frac{(V-\a I)}{U-\l} \Omega ({\bf u},\l) 
\lab{vulp}\\
\pa_i \Omega ({\bf u},\l) &=& \(V_i + \frac{E_{ii}(V-\a I)}{\l-u_i}\)  \Omega ({\bf u},\l) \,. 
\lab{uiphi}
\er
We will now show that $\Omega ({\bf u},\l)$ defined as :
\be
\Omega ({\bf u},\l)  = {\rm Res}_{z} \( \Psi ({\bf u}, z) z^{-1-\a} e^{-\l\, z}\)
\lab{defphi}
\ee
with the wave function $\Psi ({\bf u}, z)$ which satisfies
equations \rf{djpsi}-\rf{vironpsi}, 
will satisfy eqns. \rf{vulp}-\rf{uiphi}.

It follows that :
\br
&&(U-\l I)\frac{\partial\Omega({\bf u},\l)}{\partial \l}
=-{\rm Res}_{z} (U-\l I) z^{-\a} \Psi({\bf u},z) e^{-\lambda z}\nonu \\
&=&-{\rm Res}_{z} \(U\Psi({\bf u},z) z^{-\a} e^{-\lambda z}+\Psi({\bf u},z)
z^{-\a} \frac{\partial e^{-\lambda z}}{\partial z}\) \nonu \\
&=&-{\rm Res}_{z} \(U\Psi({\bf u},z) z^{-\a} e^{-\lambda z}-
\frac{\partial \Psi({ {\bf u}},z)}{\partial z}z^{-\a} e^{-\lambda z}
+\a z^{-1-\a} \Psi({ {\bf u}},z) e^{-\lambda z}\)\nonu\\
&=&{\rm Res}_{z} (V-\a I) ({\bf u}) z^{-1-\a}\Psi({ {\bf u}},z)e^{-\lambda z}
=(V-\a I) ({\bf u})\Omega({\bf u},\l) \,. \lab{vulp-proof}
\er
Equation \rf{vironpsi} was used in this derivation. 
Hence $\Omega ({\bf u}, \l) $
indeed satisfies eq. \rf{vulp}.
Furthermore, due to equation \rf{djpsi} it also holds that
\br
\pa_i \Omega ({\bf u},\l) &=&   V_i \Omega ({\bf u},\l) +E_{ii} 
{\rm Res}_{z} \(z^{-\a}  \Psi ({\bf u}, z) e^{-\l\, z}\) =
 V_i \Omega ({\bf u},\l) - E_{ii} \frac{\pa}{\pa \l} \Omega ({\bf u},\l)\nonu \\
&=&   V_i \Omega ({\bf u},\l)+ E_{ii}\frac{(V-\a I)}{\l-U} \Omega ({\bf u},\l) 
\lab{paiPz}
\er
which agrees with \rf{uiphi}.
Hence $W$ from eqs.\rf{wone} is given by $W ({\bf u}, \l)
= M \Omega ({\bf u}, \l)$.

Note, that due to \rf{djpsi}-\rf{vironpsi} $\Psi ({\bf u}, z)$ satisfies
$(z d /d z - \sum_{i=1} u_i \pa_i) \Psi ({\bf u}, z)=0$.
This leads to equation
$(\l d /d \l + \sum_{i=1} u_i \pa_i) \Omega ({\bf u}, \l)=0$.

Plugging $S_i$ into \rf{tau} we get
\be
\pa_j \log \tau = 
\sum_{k=1,\ k\neq j}^N  \frac{\tr(S_j S_k)}{u_j-u_k}
= \sum_{k=1,\ k\neq j}^N  \frac{\tr(E_{jj} V E_{kk}V)}{u_j-u_k}
=\h \tr (V_j V)
\lab{firtau}
\ee
which reproduces the well-known result for the isomonodromic tau
function \rf{tauvjv}.
The isomonodromic tau function $\tau$  is related to 
Dubrovin's \ct{Du-lectures} isomonodromic tau function 
$\tau_I$ as follows: $\tau_I=1/\sqrt{ \tau}$ \ct{LM,needs-paper}.

\section{Darboux-Egoroff Metric, the Two-dimensional Case}
\label{ntwo}
For $N=2$ there are only two canonical coordinates from which one can
construct function $\tau_0 = u_1-u_2$ such that $I(\tau_0)=0,
E(\tau_0)=\tau_0$. Then the  tau function
$\tau= \tau_0^{R^2}$
satisfies $I(\tau)=0,
E(\tau)=R^2 \tau$.
In order to satisfy equation \rf{djpsi} we take
$\b_{21}=\b_{12}= i R /\tau_0$ and we find in terms of the Pauli matrices :
\be
V_j = \sbr{\b}{E_{jj}} = \pa_j (R \log \tau_0 \sigma_2)
\qquad V= \sbr{\b}{U}  =R \sigma_2\, .
\lab{vst}
\ee
Solution to equation $(\pa_j - V_j) M_0 =0$ can be calculated explicitly
in $N=2$ and is given by  $ M_0 = \exp \(\sigma_2  R \log \tau_0  \)$.
Let
\be
S = \frac{1}{\sqrt{2}} \fourmat{-1}{-1}{-i}{i} , \quad
S^{-1} = \frac{1}{\sqrt{2}} \fourmat{-1}{i}{-1}{-i},
\;\; \eta = S^T S = \fourmat{0}{1}{1}{0}\, .
\lab{ssinv}
\ee
Then $ \mu = M^{-1} V M =R \sigma_3$ for $ M = M_0 S$.
Also,
\be
\cU = M^{-1} U M =\h  \fourmat{u_1+u_2}{\tau_0^{1-2R}}
{\tau_0^{1+2R}}{u_1+u_2} 
\lab{ntcu}
\ee
and
\be
\pa_1 C = M^{-1} E_{11} M = \h \fourmat{1}{\tau_0^{-2R}}{\tau_0^{2R}}{1}
, \;\;\;
\pa_2 C = M^{-1} E_{22} M = 
\h \fourmat{1}{-\tau_0^{-2R}}{-\tau_0^{2R}}{1}\, .
\lab{paotcc}
\ee

For the matrix $\Xi^{(n)}$ we have :
\be
E\, (\Xi^{(n)}) = \(n I - {\rm ad}_{\mu} \) \Xi^{(n)} = 
\fourmat{n\Xi^{(n)}_{11}}{(n-2R)\Xi^{(n)}_{12}}{(n+2R)\Xi^{(n)}_{21}}
{n\Xi^{(n)}_{22}} = \cU \Xi^{(n-1)}
\lab{ezxinnt}
\ee
and therefore from $E (\Xi^{(1)})= E (C)= (I- {\rm ad}_{\mu})C=\cU $ and
\rf{ntcu} 
we derive
\be
C = \h \fourmat{u_1+u_2}{\frac{\tau_0^{1-2R}}{1-2R}}
{\frac{\tau_0^{1+2R}}{1+2R}}{u_1+u_2}=
\fourmat{x^1}{\frac{1}{2(1-2R)}(2(1+2R)x^2)^{\frac{1-2R}{1+2R}}}
{x^2}{x^1},
\lab{cmnh}
\ee
valid for $ R \ne \pm \h$.
Note, that $E= x^1 \pder{}{x^1}+ (1+2R) x^2 \pder{}{x^2}$
and
\be
\cU = 
\fourmat{x^1}{\frac{1}{2} (2(1+2R)x^2)^{\frac{1-2R}{1+2R}}}
{(1+2R) x^2}{x^1}\, .
\lab{unh}
\ee

Using expression \rf{ssinv} we find :
\be
\p^{(2)}_i =  \Xi^{(2)}_{2\,i} ,\, i=1,2 \quad \;
\p^{(3)}_1 =  \Xi^{(3)}_{2\,1}
\lab{pptt}
\ee
which when plugged into expression  \rf{fep} or
 $2 F = - \p_{1}^{(3)} + 
 \sum_\d x^\d \p_{\d}^{(2)} $ yields
\be
2 F =  -\Xi^{(3)}_{2\,1} + x^1 \Xi^{(2)}_{2\,1} +x^2 \Xi^{(2)}_{2\,2}\,.
\lab{fept}
\ee
We will use equation \rf{ezxinnt}, which for $n=2$ reads
\be
\(2 I - {\rm ad}_{\mu} \) \Xi^{(2)} = 
\fourmat{2\Xi^{(2)}_{11}}{(2-2R)\Xi^{(2)}_{12}}{(2+2R)\Xi^{(2)}_{21}}
{2\Xi^{(2)}_{22}} = \cU C
\lab{xint}
\ee
and plugging $\cU$ from \rf{ntcu} into the above relation
yields
\br
\Xi^{(2)}_{1\,1} &= & \frac{1}{8} \( (u_1+u_2)^2
+ \frac{\tau_0^2}{1+2R} \) = \h (x^1)^2 + 
\frac{(2(1+2R)x^2)^{\frac{2}{2R+1}}}
{8(1+2R)} \lab{xi2oo}\\
\Xi^{(2)}_{2\,2} &= &  \frac{1}{8} \( (u_1+u_2)^2
+ \frac{\tau_0^2}{1-2R} \)=\h (x^1)^2 + 
\frac{(2(1+2R)x^2)^{\frac{2}{2R+1}}}{8(1-2R)}
 \lab{xi2tt}\\
\Xi^{(2)}_{21}&= &  \frac{1}{4 (2+2R)} \( (u_1+u_2) \tau_0^{1+2R}
(\frac{1}{1+2R} +1) \) = x^1 x^2  \lab{xi2to}
\er
where we used identification between $x^1, x^2 $
and the first column of the $C$ matrix in \rf{cmnh}.
Furthermore, for \rf{ezxinnt} with $n=3$ we find
\be
\Xi^{(3)}_{2\,1} =  \frac{1}{2(3+2R)}\( \tau_0^{2R+1} 
\Xi^{(2)}_{1\,1} +(u_1+u_2) \Xi^{(2)}_{2\,1} \) 
= \frac{1}{2(3+2R)}\( 2(1+2R)x^2 \Xi^{(2)}_{1\,1}
+ 2x^1 \Xi^{(2)}_{2\,1} \) 
\lab{xitto}
\ee
or 
\be
\Xi^{(3)}_{2\,1} =  \frac{1}{2(3+2R)}\( (1+2R)(x^1)^2 x^2 
+2(x^1)^2 x^2 + \frac{(2(1+2R)x^2)^{\frac{3+2R}{2R+1}}}
{8(1+2R)} \)\, .
\lab{xitth}
\ee
Plugging it into \rf{fept} gives
\be
F = \h (x^1)^2 x^2 + \frac{(2(1+2R)x^2)^{\frac{3+2R}{2R+1}}}
{16(3+2R) (1-2R)}
\lab{tfsa}
\ee
valid for $R\ne -3/2, R \ne \pm 1/2$.
The remaining special cases of $R= -3/2, \pm 1/2$
must be considered separately.

\sskp
{\bf R=-3/2 }
\sskp
The problem arises for $R=-3/2$ due to the fact that
the matrix operator $\(n I - {\rm ad}_{\mu} \)$ in equation
\rf{ezxinnt} does not have an inverse for $n=3$.
So, instead of using the matrix equation \rf{ezxinnt}
we will use $E (\Xi^{(3)}) = \cU \Xi^{(2)}$ with the
Euler operator $E= 
x^1 \pder{}{x^1}+ (1+2R) x^2 \pder{}{x^2}$ being equal
for $R=-3/2$ with :
\be
E= x^1 \pder{}{ x^1}-2x^2  \pder{}{x^2}
\lab{eulrmth}
\ee
Recall from relation \rf{fept} that in order to calculate 
the superpotential $F$ we need to find the matrix element
$\Xi^{(3)}_{21}$. The relevant recursion relation is
\be
E (\Xi^{(3)}_{21}) = (\cU \Xi^{(2)})_{21}= 
-2 x^2 \Xi^{(2)}_{11} +  x^1 \Xi^{(2)}_{21} 
= - \frac{1}{2^5}
\lab{fr32}
\ee
where in the product $(\cU \Xi^{(2)})_{21}$ we used 
$\cU$, $\Xi^{(2)}_{11}$ and $ \Xi^{(2)}_{21}$ as given in 
equations \rf{unh}, \rf{xi2oo} and \rf{xi2to} with $R=-3/2$.
Solution to the differential equation \rf{fr32}
is given by
\be
\Xi^{(3)}_{21} = \h (x^1)^2 x^2 + 
\frac{1}{2^6} \log (x^2) \, .
\lab{exitrmth}
\ee
Note, that the first term on the right hand side is annihilated
by the Euler vector field \rf{eulrmth} $E\( (x^1)^2 x^2\)=0$
and therefore it can not be obtained from the relation
\rf{fr32} alone.
To obtain this term we used as additional information 
relation \rf{sumjxi} which, in view of \rf{jaci}, implies
$\pa ( \Xi^{(3)}_{21})/\pa x^1=
 \Xi^{(2)}_{21}$.
 
The prepotential according to \rf{fept} is then  :
\be
2 F =  -\Xi^{(3)}_{2\,1} + x^1 \Xi^{(2)}_{2\,1} +x^2 \Xi^{(2)}_{2\,2}
= (x^1)^2 x^2 - \frac{1}{2^6}  \log (x^2) - \frac{1}{2^7} 
, \;\;\; R = - \frac{3}{2}\, .
\lab{feptrmth}
\ee
The last term being a constant can be droped.

\sskp
{\bf R=1/2 }
\sskp
{}From \rf{paotcc} we derive
\be
C = \h \fourmat{u_1+u_2}{\log \tau_0}
{\frac{\tau_0^{2}}{2}}{u_1+u_2}=
\fourmat{x^1}{\frac{1}{2} \log (2 \sqrt{x^2})}
{x^2}{x^1},\; R =  \h
\lab{rhc}
\ee
or, since $C$ is defined up to a constant :
\be
C= \fourmat{x^1}{\frac{1}{4} \log (x^2)}
{x^2}{x^1},\; R =  \h
\lab{rhcnew}
\ee
where $x^1 = \h (u_1+u_2), \, x^2= \tau_0^{2}/4$ and therefore
the Euler operator is 
\be
E= x^1 \pder{}{ x^1}+2
x^2 \pder{}{x^2} \, .
\lab{eulrh}
\ee
Equation \rf{ntcu} gives in this case
\be
\cU = \h  \fourmat{u_1+u_2}{1}
{\tau_0^{2}}{u_1+u_2} =
\fourmat{x^1}{\frac{1}{2}}{2x^2}{x^1}=E(C)
\lab{rhntcu}
\ee
{}From $E(\Xi^{(2)})={\cal U}C$ and \rf{eulrh} we find :
\be
\Xi^{(2)} = \fourmat{\h (x^1)^2+\frac{1}{4} x^2}{\frac{x^1}{4} 
\log ({x^2})}{x^1 x^2}{\frac{(x^1)^2}{2} + \frac{x^2 }{4}
\log ({x^2}) -\frac{x^2 }{4}} \, .
\lab{exi2}
\ee
Next
\br
E (\Xi^{(3)}_{21})& =& (\cU \Xi^{(2)})_{21}= 2 x^2 \Xi^{(2)}_{11} 
+ x^1 \Xi^{(2)}_{21} = 2 (x^1)^2 x^2 +\h (x^2)^2 \nonu\\
& \to& 
\Xi^{(3)}_{21} = \h (x^1)^2 x^2 + \frac{1}{8} (x^2)^2
\lab{exith}
\er
and the prepotential according to \rf{fept} is  
\be
2 F =  -\Xi^{(3)}_{2\,1} + x^1 \Xi^{(2)}_{2\,1} +x^2 \Xi^{(2)}_{2\,2}
= (x^1)^2 x^2 +  \frac{(x^2)^2 }{4} \(\log ({x^2})-
\frac{3 }{2}\) , \;\;\; R =  \frac{1}{2}\,. 
\lab{feptrh}
\ee
\sskp
{\bf R=-1/2 }
\sskp
{}From \rf{paotcc} we derive
\be
C = \h \fourmat{u_1+u_2}
{\frac{\tau_0^{2}}{2}}{\log \tau_0}{u_1+u_2}=
\fourmat{x^1}{\frac{1}{4} e^{4 x^2}}
{x^2}{x^1},\; R =  - \h
\lab{crmh}
\ee
where $x^1 = \h (u_1+u_2), \, x^2= \h \log \tau_0$ and therefore
the Euler operator is :
\be
E= x^1 \pder{}{ x^1}+\h  \pder{}{x^2}\, .
\lab{eulrmh}
\ee
Equation \rf{ntcu} gives in this case
\be
\cU = \h  \fourmat{u_1+u_2}{\tau_0^2}
{1}{u_1+u_2} =\fourmat{x^1}{\frac{1}{2} e^{4 x^2}}{\frac{1}{2}}{x^1}
= E (C)\, .
\lab{rmhntcu}
\ee
As in the case $R=1/2$ we can determine $\Xi^{(2)}$ from 
$E(\Xi^{(2)})={\cal U}C$ and equations \rf{crmh}-\rf{rmhntcu} and 
$\Xi_{21}^{(3)}$ from $E(\Xi_{21}^{(3)})=({\cal 
U}\Xi^{(2)})_{21}$. This leads according to \rf{fept} to
\be
2 F =  -\Xi^{(3)}_{2\,1} + x^1 \Xi^{(2)}_{2\,1} +x^2 \Xi^{(2)}_{2\,2}
= (x^1)^2 x^2 + \frac{1}{32} e^{4 x^2} =
(x^1)^2 x^2 + \frac{1}{2^5} e^{4 x^2}
, \;\;\; R = - \frac{1}{2}
\lab{feptrmh}
\ee

\section{Darboux-Egoroff Metric, the Three-dimensional Case}
\label{nthree}

Let us now consider the three-dimensional manifolds.
In this case, we can rewrite the antisymmetric matrix $V$ as:
\be
V= \threemat{0}{\om_3}{-\om_2}{-\om_3}{0}{\om_1}{\om_2}{-\om_1}{0}
\lab{vome}
\ee
or $(V)_{ij}= (u_j-u_i) \beta_{ij} =\eps_{ijk} \om_k$.
{}From \rf{ebib} and \rf{ilogtv} we see
that $\omega_k$ vanishes when acted on 
by the vectorfields $E$ and $I$. That  makes  $\omega_k$ 
effectively a function of one variable $s$ such that 
$E (s)= I (s)=0$.
Let us choose 
\be
s= \frac{u_2-u_1}{u_3-u_1}\,. 
\lab{sdef}
\ee
Then equation \rf{pjv} takes a form equivalent to
the Euler top equations:
\be
\frac{d \om_1}{d s}= \frac{\om_2\om_3}{s}, \;\; \;\;\;\;
\frac{d \om_2}{d s}=  \frac{\om_1\om_3}{s(s-1)}, \;\;\;\;\;\;
\frac{d \om_3}{d s}= \frac{\om_1\om_2}{1-s} \, .
\lab{euta}
\ee
One verifies that $ d (\sum_{k=1}^3 \om^2_k)/d s =0$.
Consequently, 
\be
 \sum_{k=1}^3 \om^2_k = - R^2
\lab{rconst}
\ee
where a constant $R^2$ is an integral of equations \rf{euta}.
The same constant $R^2$ characterizes the 
homogeneity of the tau function.
Indeed, starting from expression \rf{tauvjv} one finds
for the scaling dimension \ct{Dubrovin:2001}
\be
E (\log \tau) = \h \sum_{j=1}^3 u_j\tr \( V_j V\)
= \h \tr \(  V^2\) = \h \tr \(  \mu^2\) = \h \sum_{\a=1}^3 \mu_\a^2\,. 
\lab{wtau}
\ee
Recalling that $(V)_{ij}= \eps_{ijk} \om_k$
we can rewrite the above as :
\be
E (\log \tau) = \h \sum_{j=1}^3 \sum_{i=1}^3 (\eps_{ijk} \om_k)^2
= - \sum_{k=1}^3 \om^2_k = R^2 \, .
\lab{etaur}
\ee
As shown in \ct{agz02}, for $\eta_{11}$ different from zero the
homogeneity of the Lam\'e coefficients $h_i$ must vanish.
In such case, the Lam\'e coefficients $h_i$ depend only on one
variable $s$ due
to the fact that $I(h_i)=E(h_i)=0$.
The relations $\pa_j h_i^2 = \pa_i h_j^2$ translate for the function
$h_i^2(s)$ to
\be
s \dder{h_1^2}{s} =(s-1)s \dder{h_2^2}{s} =(1-s) \dder{h_3^2}{s} \,. 
\lab{hofs}
\ee
Also, since
\be
\om_k = \frac{u_j-u_i}{2 h_ih_j} \pder{h_i^2}{u_j}
, \quad i,j,k \;\mbox{cyclic}
\lab{omkhij}
\ee
we find e.g.
\be
\om_3 =  \frac{s}{2h_1h_2} \dder{h_1^2}{s}, \;\;
\om_2 = \frac{s}{2h_1h_3} \dder{h_1^2}{s}
\ee
and so $h_3 \om_2=h_2 \om_3$ and similarly $h_1 \om_2=h_2 \om_1$.
We conclude that
\be
\om_i^2 = -\frac{R^2}{\eta_{11}} h_i^2, \quad i=1,2,3
\lab{omihi}
\ee
and comparing equations \rf{euta} with equation \rf{hofs}
we obtain like in \ct{segert2} :
\be
s \dder{h_1^2}{s} =(s-1)s \dder{h_2^2}{s} =(1-s) \dder{h_3^2}{s} 
= -2 i \frac{R}{\sqrt{\eta_{11}}} h_1 h_2 h_3\,. 
\lab{hofsa}
\ee

\section{Rational Landau - Ginsburg Models }
\label{lg-models}
In this section we will show how to associate the canonical
Darboux-Egoroff structure to the rational Landau - Ginsburg models.
Following Aoyama and Kodama \ct{aoyama} we study a rational potential :
\br
W (z) &=& \frac{1}{n+1} z^{n+1} +a_{n-1} z^{n-1}+ {\ldots} +a_0
+ \frac{v_1}{z-v_{m+1}}+ \frac{v_2}{2(z-v_{m+1})^2} + {\ldots} 
\nonu\\
&+&
\frac{v_m}{m(z-v_{m+1})^m}
\lab{ratW}
\er
which is known to characterize
the topological Landau-Ginzburg (LG) theory.
The rational potential in this form can be regarded as the Lax
operator of a particular reduction of the dispersionless 
KP hierarchy \ct{aoyama,Chang:2001qz,strachan}

The space of rational potentials from
 \rf{ratW} is naturally endowed with the metric :
\be
g ( \pa_t W, \pa_{t^{\pr}} W) = {\rm Res}_{z \in {\rm Ker} W^{\pr}}
\( \frac{ \pa_t W \pa_{t^{\pr}} W}{W^{\pr}} \)dz
\lab{wmetric}
\ee
where $\pa_t W = \pa_t a_{n-1} z^{n-1}+ {\ldots} + \pa_t a_0
+ \frac{\pa_t v_1}{z-v_{m+1}} +{\ldots} $
describes a tangent vector to the space of rational potentials obtained by
taking derivative of all coefficients with respect to their argument.
${W^{\pr}} (z)$ is a derivative with respect to $z$
of the  rational potential $W$ :
\be
{W^{\pr}} (z)=  z^{n} +(n-1) a_{n-1} z^{n-2}+ {\ldots} 
- \frac{v_m}{(z-v_{m+1})^{m+1}} \,. 
\lab{wprime}
\ee
Next, we find the flat coordinates $x_\a, \a=1,{\ldots} ,m+1$
and ${\ti x}_\g, \g =1 ,..,n$ such that 
\be
g (\frac{\pa W}{ \pa x_\a}, \frac{\pa W }{\pa x_\b})
= \eta_{\a\b}, \;\;
g (\frac{\pa W }{ \pa {\ti x}_\g}, \frac{\pa W }{\pa {\ti x}_\d})
= {\ti \eta}_{\g\d},\;\;
g (\frac{\pa W }{ \pa x_\a}, \frac{\pa W }{\pa {\ti x}_\g})
= 0
\ee
with constant and non-degenerate matrices $\eta_{\a\b}$ and ${\ti \eta}_{\g\d}$.

Consider first the function $w=w(W,z)$ such that $W(z) = w^{-m}/m$ 
and $z = x_{m+1} +x_{m} w + {\ldots} + x_1 w^m=
\sum_{\a=1}^{m+1}x_\a w^{m+1-\a}$. We take $z \sim x_{m+1}$ or
$\v w \v \ll 1$. It follows that
\be
W^{\pr} dz = - \frac{1}{w^{m+1} } dw, \;\;\;\;\;\;
\frac{\pa W}{ \pa x_\a} = W^{\pr} \frac{\pa z}{ \pa x_\a} 
= W^{\pr} w^{m+1-\a}
\ee
Consequently:
\br
&&g (\frac{\pa W}{ \pa x_\a} , \frac{\pa W}{ \pa x_\b})= 
 - {\rm Res}_{z=\infty}\( \frac{(\pa W / \pa x_\a) (\pa W /\pa x_\b)}
{W^{\pr}}\) dz \lab{metrx} \\
&=& 
- {\rm Res}_{z=\infty} \( W^{\pr} w^{m+1-\a}w^{m+1-\b}\) dz 
= 
{\rm Res}_{w=\infty} \(\frac{w^{m+1-\a}w^{m+1-\b}}{w^{m+1} }\) dw=
 \d_{\a+\b=m+2} \,. 
\nonu 
\er
Hence $x_\a$ are flat coordinates with the metric $\eta_{\a\b}=
\d_{\a+\b=m+2} $. The coefficients $v_j,\,j=1,{\ldots}, 
m+1$ of $W(z)$ are given in terms of the flat coordinates as 
\ct{aoyama}:
\br
v_k &=& \sum_{\a_1+{\ldots} +\a_k=(k-1)m+k} x_{\a_1}x_{\a_2}\cdots
x_{\a_k}, \quad k=1,{\ldots} ,m 
\lab{vkasx}\\
v_{m+1} &=& x_{m+1}\,.  \nonu
\er
Examples are :
\be
v_m = (x_m)^m, \; v_{m-1} = (m-1) x_{m-1}(x_m)^{m-2}, \,
{\ldots}, \, v_1=x_1\,. 
\lab{vkasxe}
\ee
To represent the remaining coefficients of $a_i, i=1,{\ldots} ,n$ 
of $W$ in terms of the flat coordinates we consider
a relation:
\be
z = w+ \frac{{\ti x}_1}{w} + \frac{{\ti x}_2}{w^2}+{\ldots} +
\frac{{\ti x}_n}{w^n}
\lab{anxti}
\ee
valid for large $z$ and $\v w \v  \gg 1$. In this limit we impose a relation
$W= w^{n+1}/(n+1)$ from which it follows that
\be
W^{\pr} dz= w^n dw
, \;\;\quad 
\frac{\pa W}{ \pa {\ti x}_\g} = W^{\pr} \frac{\pa z}{ \pa {\ti x}_\g} 
= W^{\pr} w^{-\g}\,. 
\lab{Wprwn}
\ee
We find 
\br
&&g (\frac{\pa W}{ \pa {\ti x}_\g} , \frac{\pa W}{ \pa {\ti x}_\d})= 
{\rm Res}_{z\in {\rm Ker} W}\( \frac{(\pa W / \pa {\ti x}_\g )
(\pa W /\pa {\ti x}_\d)}
{W^{\pr}}\) dz  
\lab{metrxa}\\
&=&  {\rm Res}_{z \in {\rm Ker} W} 
\( W^{\pr} w^{-\g}w^{-\d}\) dz =
{\rm Res}_{w=0} w^{n-\g-\d} dw=
 \d_{\g+\d=n+1} \,. 
\nonu
\er
Hence ${\ti x}_\g$ are flat coordinates with the metric 
${\ti \eta}_{\g\d}=
\d_{\g+\d=n+1} $.
By similar considerations $\eta_{\a \g}=0$ for $\a=1,{\ldots} ,m+1, \g =1,{\ldots}
,n$.

{}From expression \rf{anxti} and $W(z) = w^{n+1}/(n+1)$ one can find
relations between coefficients $a_\g$ and ${\ti x}_\g$ \ct{aoyama}
starting with $a_{n-1} = -{\ti x}_1$ and so on.

We will now show how to associate to the rational potentials $W$ 
canonical coordinates $u_i, i=1,{\ldots} ,n+m+1$ for which
the metric \rf{wmetric} becomes a Darboux-Egoroff metric.

Let $\a_i$, $i=1,{\ldots} ,n+m+1$ be roots of the rational potential 
$W(z)$ in \rf{wprime}. Equivalently,
$W^{\pr}(\a_i) =0$ for all $i=1,{\ldots} ,n+m+1$.
Thus $W^{\pr} (z)$ can be rewritten as 
\be
W^{\pr} (z) = \frac{\prod_{j=1}^{n+m+1}(z-\a_j)}{(z-v_{m+1})^{m+1}}\,. 
\lab{wprima}
\ee
Next, define the canonical coordinates as 
\be
u_i = W (\a_i) , \quad i=1,{\ldots} ,n+m+1\,. 
\lab{cancoord}
\ee
The identity :
\br
\d^i_j &=& \frac{\pa u_i}{\pa u_j} =  \frac{\pa  W (\a_i)}{\pa u_j} 
\nonu \\
&=& W^{\pr} (\a_i) \frac{\pa \a_i}{\pa u_j} + \frac{\pa  W }{\pa u_j} (\a_i)
= \frac{\pa  W }{\pa u_j} (\a_i)
\lab{wuiuj}
\er
implies that 
\be
\frac{\pa  W }{\pa u_j} (z) =  
\frac{\pa a_{n-1}}{\pa u_j} z^{n-1}+ {\ldots} +
\frac{\pa a_{0}}{\pa u_j}
+ \frac{\pa v_1/ \pa u_j}{z-v_{m+1}}+ {\ldots} 
+
\frac{v_m}{(z-v_{m+1})^{m+1}} \frac{\pa v_{m+1}}{\pa u_j}
\lab{wuja}
\ee
can be rewritten as 
\be
\frac{\pa  W }{\pa u_j} (z) =  
\frac{\prod_{k=1,j\ne k}^{n+m+1}(z-\a_k)}{(z-v_{m+1})^{m+1}}
\, \frac{(\a_j-v_{m+1})^{m+1}}{\prod_{k=1,j\ne k}^{n+m+1}(\a_j-\a_k)}\,. 
\lab{wujb}
\ee
Consider
\be
g ( \frac{\pa  W }{\pa u_i}, \frac{\pa  W }{\pa u_j}) = 
{\rm Res}_{z \in {\rm Ker} W^{\pr}} 
\( \frac{ ({\pa  W }/{\pa u_i})({\pa  W }{\pa u_j})}{W^{\pr}} \)dz\,. 
\lab{wmetuu}
\ee
Recalling \rf{wprima} and \rf{wujb} we find that 
$g ( {\pa  W }/{\pa u_i}, {\pa  W }/{\pa u_j})=0$ for $i\ne j$.
For $i=j$, we find
\br
g ( \frac{\pa  W }{\pa u_i}, \frac{\pa  W }{\pa u_i}) &= &
{\rm Res}_{z \in {\rm Ker} W^{\pr}} 
\( \frac{( {\pa  W }/{\pa u_i})^2}{W^{\pr}} \)dz \nonu \\
&=& 
\frac{(\a_i-v_{m+1})^{m+1}}{\prod_{j=1,j\ne i}^{n+m+1}(\a_i-\a_j)}
= \frac{ \pa a_{n-1}}{\pa u_i}
\lab{wmetuii}
\er
where the last identity was obtained by comparing coefficients of
the $z^{n-1}$ term in \rf{wuja} and \rf{wujb}.

Hence, in terms of the coordinates $u_i$ the metric can be rewritten
as $ g = \sum_{i=1}^N h_i^2(u) (d u_i)^2$ with the Lam\'e
coefficients :
\be
 h_i^2(u) = \frac{ \pa a_{n-1}}{\pa u_i}\,. 
\lab{lamea}
\ee

\subsection{N=3 Model, Example of Rational Landau - Ginsburg models }
\label{examples}
Consider the model with $n=m=1$ in \rf{ratW}:
\be
W (z) = \h z^2 +x_1 +\frac{x_2}{z-x_3}
\lab{oldmodel}
\ee
where as coefficients we used the flat coordinates $x_1 =-{\ti x}_1$
and $x_2,x_3$ corresponding to $x_1,x_2$ of the previous section.
The flat coordinates $x_\a , \a=1,2,3$ 
are related to the flat metric :
\be
\eta^{\a\b} = \eta_{\a\b} = {\rm Res}_{z \in {\rm Ker} W^{\pr}}
\( \frac{ (\pa W /\pa x_\a)( \pa W /\pa x_\b) }{W^{\pr}} \)dz
= \threemat{1}{0}{0}{0}{0}{1}{0}{1}{0}\,. 
\lab{wmetric3}
\ee
The metric tensor can be derived from the more general expression involving
the structure constants
\be
c^{\a\b\g} = {\rm Res}_{z \in {\rm Ker} W^{\pr}}
\( \frac{ (\pa W /\pa x_\a)( \pa W /\pa x_\b)( \pa W /\pa x_\g) }{W^{\pr}} \)dz
\lab{strco}
\ee
through relation $\eta^{\a\b}= c^{\a\b1}$.
The non-zero values of the components of $c_{\a\b\g} $ are found from 
\rf{strco} to be :
\be
c^{111} = 1 ,\; c^{123}=1 , c^{222}= 1/x_2, \; c^{233}=x_3
, \; c^{333}=x_2
\ee
the other values can be derived using that $c^{\a\b\g} $ is symmetric in all
three indices.
These values can be reproduced from the formula \rf{cfxx}
with the prepotential :
\be
F (x_1, x_2, x_3) = \frac{1}{6} x_2 (x_3)^3 +  
\frac{1}{6} (x_1)^3 + x_1  x_2 x_3 +
\h (x_2)^2 \( \log x_2 -  \frac{3}{2}\)\,. 
\lab{prepo3}
\ee

The prepotential satisfies the quasi-homogeneity relation 
\rf{quasi} with $d_F=3$ with respect to the 
Euler vectorfield :
\be
E= x_1 \frac{\pa }{\pa x_1} + \frac{3}{2} x_2 \frac{\pa }{\pa x_2} +
\frac{1}{2}x_3 \frac{\pa }{\pa x_3} 
= x^1 \frac{\pa }{\pa x^1} + \frac{1}{2} x^2 \frac{\pa }{\pa x^2} +
\frac{3}{2}x^3 \frac{\pa }{\pa x^3} \,. 
\lab{euler3}
\ee
We now adopt a general discussion of canonical coordinates from Section \ref{nthree}
to the case of $n=m=1$. 
Let $\a_i$, $i=1,2,3$ be roots of the polynomial 
$W^{\pr} (z) = z -{x_2}/{(z-x_3)^2}$.
So, $\a_i$ satisfy $W^{\pr}(\a_i) =0$ or $\a_i (\a_i -x_3)^2-x_2=0$
for all $i=1,2 ,3$.

Then, it follows by taking derivatives of $\a_i (\a_i -x_3)^2=x_2$
with respect to $x_2,x_3$ that 
\be
\frac{\pa \a_i}{\pa x_3} = \frac{2 \a_i}{3\a_i-x_3},\;\;\qquad 
\frac{\pa \a_i}{\pa x_2} = \frac{1}{(\a_i-x_3)(3\a_i-x_3)}
\lab{alpix}
\ee
and further that
\be
\frac{\pa u_i}{\pa x_3} = \frac{x_2 }{(\a_i-x_3)^2}=\a_i ,\;\;\qquad 
\frac{\pa u_i}{\pa x_2} = \frac{1}{\a_i-x_3}
\lab{uix}
\ee
for the canonical coordinates $u_i = W(\a_i)=\h \a_i^2 + x_1 + x_2/(\a_i-x_3)$.
We now present a method of inverting the derivatives in \rf{uix} or
alternatively to find the matrix elements $m_{ij}$ of the matrix 
$M$ from relation \rf{mdef}.
The sum of the canonical coordinates is equal to 
$\sum_{i=1}^3 u_i = 3 x_1 + x_3^2$ and therefore 
\be
1 = 3 \frac{\pa x_1}{\pa u_i} + 2 x_3 \frac{\pa x_3}{\pa u_i}
=h_i^2 \(3+ 2 x_3 \frac{\pa u_i}{\pa x_2}\)
\lab{1himi3}
\ee
where we used the fact that
\be
\frac{\pa x_3}{\pa u_i} = m^2_{i1} \frac{\pa u_i}{\pa x_2}
\lab{x3uia}
\ee
because of
\be
\frac{\pa x_\a}{\pa u_i} = m_{i1} m_{i\a} , \quad
\frac{\pa u_i}{\pa x_\a} = \eta_{\a \b} \frac{m_{i\b} }{m_{i1}}, \;\;\; 
h_i^2 = m^2_{i1}
\ee
Hence, from relation \rf{1himi3} it holds that $h_i^2=  \(3+ 2 x_3 \frac{\pa u_i}{\pa x_2}\)^{-1}$
or by using equation \rf{uix} that 
\be
\frac{\pa x_1}{\pa u_i} = h_i^2=  \frac{\a_i-x_3}{3\a_i-x_3}\,. 
\lab{x1ui}
\ee
Plugging the last equation into equation \rf{x3uia} and using relation
\rf{uix} we obtain
\be
\frac{\pa x_3}{\pa u_i} = \frac{1}{3\a_i-x_3}\,. 
\lab{x3uib}
\ee
Similarly, from 
\be
\frac{\pa x_2}{\pa u_i} = m^2_{i1} \frac{\pa u_i}{\pa x_3}
\lab{x2uia}
\ee
we obtain
\be
\frac{\pa x_2}{\pa u_i} = \frac{x_2}{(\a_i-x_3)(3\a_i-x_3)}=
 \frac{\a_i(\a_i-x_3)}{(3\a_i-x_3)}\,. 
\lab{x2uib}
\ee
Furthermore,
\be
\frac{\pa \a_i}{\pa u_j}= \frac{\pa \a_i}{\pa x_2}\frac{\pa x_2}{\pa u_j}
+\frac{\pa \a_i}{\pa x_3}\frac{\pa x_3}{\pa u_j}
\ee
gives for $i\ne j$:
\be
\frac{\pa \a_i}{\pa u_j}= \frac{1}{(3\a_i-x_3)(3\a_j-x_3)}\(
\frac{\a_j(\a_j-x_3)}{(3\a_i-x_3)}+2 \a_i\)
\lab{paapauj}
\ee
and for $i = j$ :
\be
\frac{\pa \a_i}{\pa u_i}= \frac{3\a_i}{(3\a_i-x_3)^2}\,. 
\lab{paapaui}
\ee
Using \rf{paapauj} we can take a derivative of $h^2_i$ in \rf{x1ui} 
and find the rotation coefficients defined in \rf{rotco}
to be
\be
\b_{ij} = - \frac{(\a_k-x_3)(3\a_k-x_3)}{(3\a_i-x_3)(3\a_j-x_3)}
\frac{1}{\sqrt{(\a_i-x_3)(3\a_i-x_3)(\a_j-x_3)(3\a_j-x_3)}}\,. 
\lab{bmodel}
\ee
Its square is then
\be
\b_{ij}^2 = - \frac{1}{(\a_i-\a_j)^2}\frac{1}{(4x_3-3\a_k)^2}
\frac{\pa x_1}{\pa u_k}, 
\lab{bsmodel}
\ee
where $i,j,k$ are cyclic.
Recall that in equation \rf{vome} we have introduced the
functions $\om_k = (u_j-u_i) \b_{ij}$, where again we used the 
cyclic indices $i,j,k$.
The difference of canonical coordinates can be written as :
$u_j-u_i=(\a_i-\a_j)(3\a_k-4x_3)/2$ which together with equation \rf{bmodel}
yields:
\be
\om_k^2 =- \frac{1}{4} h_k^2 =- \frac{1}{4}\frac{\pa x_1}{\pa u_k}
=- \frac{1}{4} \frac{\a_k-x_3}{3\a_k-x_3}\,. 
\lab{omksmodel}
\ee
Since $ I=\sum_{i=1}^3 \pa / \pa u_i= \pa / \pa x_1$
then 
\be
\sum_{k=1}^3 \om_k = - \frac{1}{4}, \quad 
E (\log \tau) = \frac{1}{4}\,. 
\lab{etaurmd}
\ee
The explicit form of the roots $\a_i$ is needed to find
expressions for $\om_k$ and its dependence on the parameter
$s$.
It is convenient to introduce $q= x_2/(x_3)^3 $ and
$a_i = \a_i/x_3$ which 
satisfy equation $a_i(a_i-1)^2=q$.
Let us furthermore introduce a parameter $ \om$ such that
$q =4 (\om^2-1)^2 /(\om^2+3)^3$. This parametrization makes
it possible to obtain the compact expressions for $\om_k$.
The three solutions to the algebraic equation 
\be
a(a-1)^2 =q= 4 \frac{(\om^2-1)^2 }{(\om^2+3)^3}
\lab{paraeq}
\ee
are:
\be
a_1 = \frac{4 }{\om^2+3}, \;\; a_2 = \frac{(\om+1)^2 }{\om^2+3}, \;\; 
a_3 = \frac{(\om-1)^2 }{\om^2+3}\,. 
\lab{aroots}
\ee
Note, that $a_2 \leftrightarrow a_3$ under $\om \leftrightarrow -\om$
transformation, which shows that $\om$ is a purely imaginary 
variable.
First, we find that the variable $s$ from \rf{sdef} can be expressed as :
\be
s= \frac{(a_2-a_1)}{(a_3-a_1)}\frac{(3a_3-4)}{(3a_2-4)}=
\frac{(\om-3)^3(\om+1)}{(\om+3)^3(\om-1)}
\lab{sdefmodel}
\ee
Next, from relations $h_i^2= (a_i-1)/(3a_i-1)$ and equation  
\rf{omksmodel} we derive :
\be
\om_1^2 = - \frac{1}{4}\frac{(\om^2-1) }{(\om^2-9)}, \;\;\;
\om_2^2 =  \frac{1}{4}\frac{(\om+1) }{\om(\om-3)}, \;\;\;
\om_3^2 = - \frac{1}{4}\frac{(\om-1) }{\om(\om+3)}\,. 
\lab{ompara}
\ee
They provide solutions to the Euler top equations 
\rf{euta}.
The corresponding function \ct{segert1,segert2}
\be
y(\om) = \frac{( \om -3 )^2({\om} +1  ) }{( \om +3 )({\om^2}+3 )}
\lab{painsol}
\ee
connected with $\om_k$'s through relations \ct{H1,H2,H3}:
\br
\om^2_1&=&-\frac{(y-s)y^2(y-1)}{s}\left(v-\frac{1}{2(y-s)}\right)\left(
v-\frac{1}{2(y-1)}\right)~,\nonu \\
\om^2_2 &=&
\frac{(y-s)^2y(y-1)}{s(1-s)}\left(v-\frac{1}{2(y-1)}\right)\left(
v-\frac{1}{2y}\right)~,\nonu \\
\om^2_3&=&-\frac{(y-s)y(y-1)^2}{(1-s)}\left(v-\frac{1}{2y}\right)\left(
v-\frac{1}{2(y-s)}\right) 
\lab{omtoy}
\er
with the auxiliary variable $v$ defined by equation
\be
\dder{y}{s}=\frac{y(y-1)(y-s)}{s(s-1)}\left(2v-\frac{1}{2y}-\frac{1}{2(y-1)}
+\frac{1}{2(y-s)}\right)
\lab{auxi}
\ee
is a solution of the Painlev\'e VI  equation  \ct{H1,H2,H3}:
\br
\frac{d^2 y}{d s^2} &=&\frac{1}{2}\left( \frac{1}{y} 
+\frac{1}{y-1}+\frac{1}{y-s}\right)
(\dder{y}{s})^2- \left( \frac{1}{s} +\frac{1}{s-1}+\frac{1}{y-s}\right)
\dder{y}{s}\nonu \\
&  +& \frac{y(y-1)(y-s)}{s^2(s-1)^2}\left[
\frac{1}{8} -\frac{s}{8y^2}+\frac{s-1}{8(y-1)^2}
+\frac{3s(s-1)}{8(y-s)^2}\right] \, .
\lab{pain6}
\er
Introducing parameter $x=(\om-3)/(\om+3)$ one can rewrite expressions
\rf{painsol} and  \rf{sdefmodel} as :
\be
y = \frac{x^2(x+2)}{x^2+x+1}\; , \qquad s= \frac{x^3(x+2) }{2x+1}\, ,
\lab{cnh2}
\ee
which reproduces the $k=3$ Poncelet 
polygon solution of Hitchin \ct{H2,H3}.

We now proceed to calculate the underlying $\t$-function.
Our knowledge of the $\t$-function is based on equation 
\rf{tauvjv} from which we derive
that 
\be
\pa_j \log \t =  \sum_{i=1}^3 \b_{ij}^2 (u_i-u_j)\,.
\lab{pajltbu}
\ee
The identity $I (\log \t) =0$, shows that $\t=\t(x_2,x_3)$
is a function of two variables $x_2,x_3$. Furthermore, it
satisfies :
\be
E (\log \t)= \( \frac{3}{2} x_2 \frac{\pa }{\pa x_2} +
\frac{1}{2}x_3 \frac{\pa }{\pa x_3}\) \log \t= \frac{1}{4}\,. 
\lab{elogtmod}
\ee
A solution to the above equation is 
\be
\log \t= \frac{1}{4} \(\frac{1}{3} \log x_2 + \log x_3\) +
f\(\frac{1}{3} \log x_2 - \log x_3\) 
\lab{decot}
\ee
where $f(\cdot)$ is an arbitrary function of it's argument. 
In order to determine the function $f$ we use equation \rf{pajltbu}
to calculate the derivative
\be
\frac{\pa \log \t}{\pa x_3} = \sum_{j=1}^3 \frac{\pa u_j}{\pa x_3}
\pa_j \log \t = \sum_{i,j=1}^3 \a_j \b_{ij}^2 (u_i-u_j)\,. 
\lab{pajx3lt}
\ee
A calculation based on equation \rf{bsmodel} yields:
\be
 x_3 \frac{\pa }{\pa x_3} \log \t=  \frac{1}{8} \frac{1}{1-\frac{27}{4}q}
= \frac{1}{4} -
f^{\pr} \(\frac{1}{3} \log x_2 - \log x_3\) 
\lab{27q}
\ee
where the last equality was obtained by 
comparing with equation \rf{decot} (recall that $q= x_2/(x_3)^3 $).
Integration gives (ignoring an inessential integration constant) :
\be
f \(\frac{1}{3} \log x_2 - \log x_3\) =
\frac{1}{24} \( \log q + \log (-4 +27 q)\) \,. 
\lab{fqres}
\ee
Using that $x_2= q x_3^3$ we can now rewrite $\log \t$
as
\be
\log \t = \frac{1}{4} \log x_3^2 + \frac{1}{24}  \log \(q^3(-4 +27 q)\) \,. 
\lab{ltresq}
\ee
Inserting parametrization of $q$ from \rf{paraeq} 
and using relation $u_2-u_3=8 x_3^2 \om^3(\om^2+3)^{-2}$
we obtain the following expression for $\log \t$ :
\be
\log \t = \log (u_2-u_3)^{\frac{1}{4}} + \frac{1}{24}  
\log \((\om-1)^6 (\om+1)^6 (\om-3)^2 (\om+3)^2 \om^{-16}\)\,. 
\lab{ltresom}
\ee
It is easy to confirm $I (\log \t)=0$ and $E (\log \t)=1/4$
based on this expression.

\acknowledgments
H.A. was partially supported by FAPESP and NSF (PHY-9820663).
A.H.Z. and J.F.G were partially supported by CNPq.


\begin{thebibliography}{99}
\bi{witten} E. Witten, On the structure of the topological phase of
two-dimeensional gravity, \npb{340}{1990}{281}
\bi{dvv} R.~Dijkgraaf, H.~Verlinde and E.~Verlinde,
Topological strings in $D < 1$, \npb{352}{1991}{59}
\bi{Du-lectures}
B. Dubrovin, Geometry of $2{\rm D}$ topological field theories, 
Springer Lect. Notes Math. {\bf 1620} (1995) 120,  
[arXiv:\href{http://arXiv.org/abs/hep-th/9407018}{\tt hep-th/9407018}]
\bibitem{Du3a}
B. Dubrovin,
Integrable systems in topological field theory, \npb{379}{1992}{627}
\bibitem{vandeLeur:2000gk}
J.W. van de Leur,
Twisted $GL_n$ loop group orbit and solutions of the WDVV equations,
{\it Intern. Math. Research Notices} {\bf 11} (2001) 551,
[arXiv:\href{http://arXiv.org/abs/solv-int/0004021}{\tt solv-int/0004021}]
\bibitem{LM}
J.W. van de Leur  and R. Martini, The construction of Frobenius manifolds from KP
tau-functions, \cmp{205}{1999}{587},
[arXiv:\href{http://arXiv.org/abs/solv-int/9808008}{\tt solv-int/9808008}]
\bibitem{needs-paper}
H.~Aratyn and J.W. van de Leur,
Integrable structure behind WDVV equations,
contribution to the NEEDS'01 Conference in Cambridge,
to appear in {\it Theor. Math. Phys.}, 
[arXiv:\href{http://arXiv.org/abs/hep-th/0111243}{\tt hep-th/0111243}]
\bibitem{Hlectures}
N. Hitchin, Frobenius Manifolds, Notes by D. Calderbank,
NATO Adv. Sci. Inst. Ser. C Math. Phys. Sci. {\bf 488},  
``Gauge Theory and Symplectic Geometry'' (Eds.) J. Hurtubise, F. Lalonde,
pgs. 69-112, Kluwer, 1997
\bibitem{prepare}
H.~Aratyn and J.W. van de Leur, paper in preparation
\bibitem{agz01}
H. Aratyn, J.F. Gomes
and A.H. Zimerman, Multidimensional Toda equations and 
topological-anti-topological fusion, to appear in {\it J. Geom. Phys.},
[arXiv:\href{http://arXiv.org/abs/hep-th/0107056}{\tt hep-th/0107056}]
\bibitem{Aratyn:2001cj}
H.~Aratyn and J.W. van de Leur,
Solutions of the WDVV equations and integrable hierarchies of KP
type,
[arXiv:\href{http://arXiv.org/abs/hep-th/0104092}{\tt hep-th/0104092}]
\bi{AKrV}A.A. Akhmetshin, I.M. Krichever, Y.S.Volvovski, A generating
formula for solutions of
associativity equations. (Russian) {\it Uspekhi Mat. Nauk} {\bf 54} (1999), no. 2(326),
167--168, 
[arXiv:\href{http://arXiv.org/abs/hep-th/9904028}{\tt hep-th/9904028}]
\bi{sch12}
L. Schlesinger, {\sl J. Reine Angew. Math.} {\bf 141} (1912) 96-145
\bi{kyoto}
M. Jimbo and T. Miwa, {\it Physica} {\bf 4} (1981) 26;
{\it Physica} {\bf 2} (1981) 407
\bi{boalch} 
P.P. Boalch, Symplectic Geometry and Isomonodromic Deformations, PhD thesis,
Oxford University, 1999
\bibitem{Dubrovin:2001}
B.~Dubrovin and  Y.~ Zhang,
Normal forms of hierarchies of integrable PDEs, Frobenius manifolds 
and Gromov - Witten invariants, 
[arXiv:\href{http://arXiv.org/abs/math.dg/0108160}{\tt math.dg/0108160}]
\bibitem{agz02}
H. Aratyn, J.F. Gomes
and A.H. Zimerman, 
[arXiv:\href{http://arxive.org/abs/math-ph/0209022}{\tt math-ph/0209022}]
\bi{segert2}
J. Segert, Frobenius manifolds from Yang-Mills instantons, 
{\it Math. Res. Lett.} {\bf 5},  no. 3 (1998)  327, 
[arXiv:\href{http://arXiv.org/abs/dg-ga/9710031}{\tt dg-ga/9710031}]
\bi{aoyama}
S. Aoyama and Y. Kodama,
Topological Landau-Ginzburg theory with a rational potential and 
the  dispersionless KP hierarchy,
\cmp{182}{1996}{185}, 
[arXiv:\href{http://arXiv.org/abs/hep-th/9505122}{\tt hep-th/9505122}];
Topological conformal field theory with a rational W potential and the dispersionless KP hierarchy,
\mpla{9}{1994}{2481}, 
[arXiv:\href{http://arXiv.org/abs/hep-th/9404011}{\tt hep-th/9404011}]
\bibitem{Chang:2001qz}
J.~H.~Chang and M.~H.~Tu,
Topological field theory approach to the generalized Benney hierarchy,
[arXiv:\href{http://arXiv.org/abs/hep-th/0108059}{\tt hep-th/0108059}]
\bi{strachan} I.A.B. Strachan,
Degenerate Frobenius manifolds and the bi-Hamiltonian structure of 
rational Lax equations, \jmp{40}{1999}{5058}, 
[arXiv:\href{http://arXiv.org/abs/solv-int/9807004}{\tt solv-int/9807004}]
\bi{segert1}
J. Segert, Painlev\'e solutions from equivariant holomorphic 
bundles, Preprint (1996), available from {\sf www.math.missouri.edu/}
${\widetilde{}}\,\,${\sf jan/papers/painpreprint.pdf}
\bibitem{H1}
N.J. Hitchin, Twistor spaces, Einsten metrics and isomonodromic 
deformations, \jdg{42}{1995}{30}
\bibitem{H2}
N.J. Hitchin, Poncelet polygons and the Painlev\'e transcendents, 
Geometry and Analysis, Oxford University Press, Bombay, 1996, 
151--185
\bi{H3}    
N.J. Hitchin, A new family of Einstein metrics, manifolds and geometry, Symposia
Mathematica series v. 36, De Bartolomeis, Triceri, and Vesentini eds., Cambridge Uni-
versity Press.
\end{thebibliography}
\end{document}